\documentclass[fleqn,11pt]{wlscirep}
\title{How to analyze data in a factorial design? \\[-0.5em] An extensive simulation study}
\usepackage[misc]{ifsym}
\author[ ]{Maria Umlauft}
\affil[ ]{Institute of Statistics, Ulm University, Germany}
\affil[ ]{Helmholtzstr. 20, 89081 Ulm, \Letter\;maria.umlauft@alumni.uni-ulm.de	}

\usepackage{lmodern}

\usepackage{multirow}
\usepackage{subcaption}
\usepackage{bbm, bm}
\usepackage{setspace}
\usepackage{amssymb}
\usepackage{lipsum}
\usepackage{apacite}
\usepackage{morefloats}

\usepackage[running]{lineno}

\usepackage{threeparttable}
\usepackage{booktabs}

\usepackage{color}
\definecolor{dkgreen}{rgb}{0,0.6,0}
\definecolor{gray}{rgb}{0.5,0.5,0.5}
\definecolor{mauve}{rgb}{0.58,0,0.82}

\usepackage{listings}
\newcommand{\bs}[1]{\mathbf{#1}}

\DeclareMathOperator{\tr}{tr} 
\def\epsilon{\varepsilon}

\numberwithin{equation}{section}

\DeclareMathOperator{\diag}{diag} 
\DeclareMathOperator{\var}{var} 
\DeclareMathOperator{\Poi}{Poi} 
\DeclareMathOperator{\rank}{rank} 
 
\DeclareMathOperator{\Logis}{Logistic}

\newcommand{\bqan}{\begin{eqnarray}}
\newcommand{\eqan}{\end{eqnarray}}
\newtheorem{definition}{{\textsc Definition}\textsc}[section]
\newtheorem{satz}[definition]{{\textsc Theorem}\textsc}

\newcommand{\bsa}{\begin{satz}}
\newcommand{\esa}{\end{satz}}
\newcommand{\bay}{\begin{array}}
\newcommand{\eay}{\end{array}}
\newcommand{\bqa}{\begin{eqnarray*}}
\newcommand{\eqa}{\end{eqnarray*}}
\def\epsilon{\varepsilon}

\keywords{ANOVA, Factorial Designs, Nonparametrics, Ranking Methods}

\begin{abstract}
Factorial designs are frequently used in different fields of science, e.g. psychological, medical or biometric studies. Standard approaches, as the ANOVA $F$-test, make different assumptions on the distribution of the error terms, the variances 
or the sample sizes in the different groups. Because of time constraints or a lack of statistical background, many users do not check these assumptions; enhancing the risk of potentially inflated type-$I$ error rates or a substantial loss of power. 
It is the aim of the present paper, to give an overview of 
different methods without such restrictive assumptions and to identify situations in which one method is superior compared to others. In particular, after summarizing their underlying assumptions, 
the different approaches are compared within extensive simulations. To also address the current discussion about redefining the statistical significance level, we also included simulations for the 0.5\% level.

\end{abstract}
\begin{document}
\doublespacing
\maketitle

\thispagestyle{empty}

\newpage
\onehalfspacing
\section{Introduction} \label{sec:intro}

Factorial designs are very common and successfully used in the wide range of life and social science, to model and infer data. Standard inference procedures to analyze such data are typically assuming equal 
variances across the different groups, equal sample sizes 
or make distributional assumptions on the error terms (e.g. normality). For example, under the conjecture of normality and homogeneity, the ANOVA $F$-test is an exact level $\alpha$ testing procedure. Nevertheless, the method results in liberal 
or conservative decisions when the assumptions are not met \cite<see, for example,>[and reference therein]{vallejo2010analysis, deshon1996alternative, rosopa2013managing}. To rectify this issue, many different methods were proposed. 
In case of heterogeneity, for instance, 
the ANOVA-type statistic proposed by \citeA{brunner1997box} and recommended in, e.g. \citeA{vallejo2010analysis}, the generalized Welch-James test \cite<see>{johansen1980welch} or the approximate degree of freedom test \cite<see>{zhang2012approximate}
are powerful and sufficient procedures. 
Moreover, resampling techniques and especially,
permutation methods were proposed \cite<see, e.g.,>{pauly2015asymptotic, umlauft2017rank}, and show to be asymptotically exact tests for general factorial design without distributional assumptions as normality or homogeneity.

Moreover, rank-based procedures are often recommended \textit{``when the assumptions for the analysis of variance
are not tenable''}, see \citeA{bewick2004statistics}. For the one-way layout, this includes the \citeA{kruskal1952use} test as well as the van der Waerden (VDW) test, which was recently recommended by \citeA{luepsen2017comparison}. 

It is the aim of this work to study and compare such inference procedures for classical, semi- and nonparametric ANOVA designs in an extensive simulation study. Thereby, a specific assessment criterion is their type-$I$ errors for 
\begin{itemize}
 \item symmetric and skewed distributions,
 \item homo- and heteroscedastic settings as well as
 \item balanced and unbalanced designs.
\end{itemize}


In this context, we also examine the behavior of the different testing procedures when the chosen significance level is 0.5\%. This was recently proposed by a large group of researchers \cite{benjamin2017redefine} to improve the reproducibility of 
scientific research. However, it has not yet been investigated if existing procedures are able to satisfactorily keep such small levels; corresponding to the possibly difficult estimation of small quantiles.

In the next section, the underlying statistical models of all studied procedures are introduced. The different approaches are established and opposed in Section~\ref{sec:testprod} and to make practical recommendations, an extensive simulation study 
is conducted, afterwards. The results and a conclusion of this work are discussed in the Section~\ref{sec:con}. A comparison between the deviation of the prerequired significance level of 5\% and 0.5\% is given in the Appendix.

\section{Statistical Model and Hypotheses}  \label{sec:mod}

The general one-way ANOVA model including $d$ different groups is given by $N=\sum_{i=1}^d n_i$ independent random variables
\begin{linenomath}
\begin{equation}\label{equ:oneway}X_{ik}= \mu_{i}+\varepsilon_{ik} = \mu + \alpha_i + \varepsilon_{ik}, \; i=1, \ldots, d \; \text{ and } \; k=1,\ldots, n_i,\end{equation}
\end{linenomath}
where the fixed effects $\alpha_i$ sum up to zero ($\sum_{i=1}^d \alpha_i=0$) and the errors $\varepsilon_{ik}$ are assumed to be independent with mean zero and identically distributed within each group ($i=1, \ldots, d$) with existing variances $\var(\varepsilon_{i1})=\sigma_i^2 \in (0,\infty)$. In 
matrix notation, the model can be expressed by 
\begin{linenomath}
\begin{equation}\label{equ:onewaymatrix}\bs X = \bs Z \bm{\mu}+ \bm{\varepsilon},\end{equation}
\end{linenomath}
where all observations are summarized in the vector $\bs X =\left(\bs X_1', \ldots, \bs X_d'\right)'$ with $\bs X_i =\left(X_{i1}, \ldots, X_{in_i}\right)'$ for $i=1,\ldots,d$. Furthermore, $\bm{\mu}=(\mu_1, \ldots, \mu_d)'$ and 
$\bm{\varepsilon}=(\varepsilon_{11}, \varepsilon_{21},\ldots,\varepsilon_{dn_d})'$ are the vectors of group-specific means, respectively error terms and $\bs Z$ is the corresponding design matrix (i.e. $\bs Z=\oplus_{i=1}^d \bs 1_{n_i}$ for the 
$n_i$-dimensional vector $\bs 1_{n_i}$ of ones).
Here, we 
distinguish between \textit{heteroscedastic} and \textit{homoscedastic} (all $\sigma_i^2 \equiv \sigma^2 \in (0,\infty)$) situations as well as \textit{parametric} models with a specific error distribution (e.g. normality) or \textit{semi- and nonparametric} 
models.  The classical ANOVA $F$-test and other testing procedures for parametric and semiparametric models are techniques for detecting differences in the group means, e.g. for the above one-way design, the corresponding null hypothesis is given by
\begin{linenomath}
\begin{equation}\label{equ:H0mu}
 H_0^\mu : \{\mu_1 = \ldots = \mu_d\}=\{\alpha_1=\ldots=\alpha_d=0\}.
\end{equation}
\end{linenomath}
This null hypothesis can be equivalently given in matrix notation as $H_0^\mu: \bs P_d \bm{\mu}=\bs 0,$ where $\bs P_d=\bs I_d-\frac{1}{d}\bs J_d$ denotes the so-called centering matrix. Here, $\bs I_d$ defines the $d$-dimensional unit matrix and the $d \times d$-
dimensional matrix of ones is given by $\bs J_d=\bs 1_d \bs 1'_d$ with $\bs 1_d$ a $d$-dimensional vector of ones.

For handling ordinal or ordered categorical data, mean-based approaches (here:~classical parametric or semiparametric models) show their limits, since means are neither meaningful nor suitable effect measures. Thus, in a purely \textit{nonparametric} general 
factorial setting, observations are assumed to be generated from  
\begin{linenomath}
\begin{equation}\label{equ:nonparamod}
 X_{ik}\sim F_i(x), \; i=1, \ldots, d \; \text{ and } \; k=1,\ldots, n_i,
\end{equation}
\end{linenomath}
where $F_i$ denotes the distribution function of group $i=1,\ldots, d$.
In contrast to the parametric and semiparametric models, the hypotheses of interest are often formulated in terms of distribution functions \cite{kruskal1952use, akritas1994fully, akritas1997nonparametric, brunner1997box, brunner2001nonparametric, umlauft2017rank} as
\begin{linenomath}
\begin{equation}\label{equ:H0f}
 H_0^F : \{ F_1 = \ldots = F_d\}.
\end{equation}
\end{linenomath}
 Again, the hypothesis can also be given in matrix notation as $H_0^F: \bs P_d \bs F=\bs 0$, where 
$\bs F=(F_1, \ldots, F_d)'$ denotes the vector of distribution functions and $\bs P_d$ is given as above. In case of homoscedastic models, the two different ways (see Equations \ref{equ:H0mu} and \ref{equ:H0f}) of formulating the null hypotheses correspond.
In particular, assuming $\sigma_i^2 \equiv \sigma^2 \in (0, \infty)$, we obtain $H_0^\mu=H_0^F$ in case of model \eqref{equ:oneway}. However, $H_0^F$ can also be inferred for the more general model \eqref{equ:nonparamod} which neither postulates
specific moment assumptions nor metric data. Thus, $H_0^\mu \subset H_0^F$ in this case. However, under heteroscedasticity, $H_0^\mu$ and $H_0^F$ have completely different implications. While $H_0^\mu$ is only concerned with mean differences and 
allows for heteroscedasticity between groups, $H_0^F$ implies a homoscedastic setting (if variances exist). Thus, $H_0^\mu  \not\subset H_0^F$ in general.

Modeling two- or higher-way layouts in factorial designs is in general done by splitting up the index $i$ in sub-indices $i_1, i_2, \ldots$. 
For example, a two-way layout with  factor $A$ (with $a$ levels), factor $B$ (for $b$ levels) and an interaction is given by setting $d=a \cdot b$ and splitting the running index in Equation \eqref{equ:oneway} into $i=1,\ldots,a$ and $j=1,\ldots,b$ writing
$\mu_{ij}=\mu+\alpha_i+\beta_j+\gamma_{ij}$. The observations are generated from
\begin{linenomath}
\[X_{ijk}=\mu+\alpha_i+\beta_j+\gamma_{ij}+\varepsilon_{ijk}, \; i=1, \ldots, a, \; j=1,\ldots,b \; \text{ and } \; k=1,\ldots, n_{ij},\]
\end{linenomath}
where again $\sum_{i=1}^a \alpha_i=\sum_{j=1}^b \beta_j=\sum_{i=1}^a\sum_{j=1}^b \gamma_{ij}= 0$ and $\varepsilon_{ijk}$ are independent random variables with mean zero and identical distributions within each factor level combination $(i,j)$ with finite variances $\sigma_{ij}^2$. 

Null hypotheses of interest are $H_0^\mu(\bs C_A)=\{\bs C_A \bs \mu=\bs 0\}=\{\left((\bs I_a-\frac{1}{a}\bs J_a)\otimes \frac{1}{b}\bs 1'_b\right)\bs \mu=\bs 0\}=\{\alpha_i\equiv 0 \; \forall i=1,\ldots,a\}$ (no main effect of factor $A$), 
$H_0^\mu(\bs C_B)=\{\bs C_B \bs \mu=\bs 0\}=\{\left(\frac{1}{a}\bs 1'_a \otimes (\bs I_b-\frac{1}{b}\bs J_b)\right)\bs \mu=\bs 0\}=\{\beta_j\equiv 0 \; \forall j=1,\ldots,b\}$ (no main effect of factor $B$) and  
$H_0^\mu(\bs C_{AB})=\{\bs C_{AB} \bs \mu=\bs 0\}=\{\left((\bs I_a-\frac{1}{a}\bs J_a) \otimes (\bs I_b-\frac{1}{b}\bs J_b)\right)\bs \mu=\bs 0\}=\{\gamma_{ij}\equiv 0 \; \forall i=1\ldots,a \text{ and } j=1,\ldots,b\}$ (no interaction effect).
In case of nonparametric designs (see Equation~\ref{equ:nonparamod}), null hypotheses are again formulated in terms of distribution functions. For an easier interpretation,  \citeA{akritas1994fully} give another representation of the null hypothesis $H_0^F$
which directly corresponds to the hypothesis formulated in terms of means as in the ANOVA setting. Thus, the group-specific distribution function $F_{ij}$ is decomposed as
\begin{linenomath}
\[F_{ij}(y)=M(y)+A_i(y)+B_j(y)+(AB)_{ij}(y),\] 
\end{linenomath}
where $\sum_{i=1}^a A_i=\sum_{j=1}^b B_j \equiv 0, \; \sum_{i=1}^a (AB)_{ij}=0 \; \forall j=1,\ldots,b \text{ and } \sum_{j=1}^b (AB)_{ij}=0 \; \forall i=1,\ldots,a$. The particular summands are than given by $A_i=\overline{F}_{i\cdot}-\overline{F}_{\cdot \cdot}=
\frac{1}{b}\sum_{j=1}^b F_{ij}-\overline{F}_{\cdot \cdot}, \; B_j=\overline{F}_{\cdot j}-\overline{F}_{\cdot \cdot}=\frac{1}{a}\sum_{i=1}^a F_{ij}-\overline{F}_{\cdot \cdot} \text{ and } (AB)_{ij}= F_{ij}-\overline{F}{i\cdot}
-\overline{F}_{\cdot j}+\overline{F}_{\cdot \cdot},$ where $\overline{F}_{\cdot \cdot}=M=\frac{1}{ab}\sum_{i=1}^a\sum_{j=1}^b F_{ij}$.
Therefore, the main effect can be tested as before with $H_0^F(\bs C_A): \{\bs C_A \bs F=\bs 0\}=\{A_i \equiv 0 \; \forall i=1,\ldots,a\}$
as the corresponding null hypothesis of main effect $A$ and with $H_0^F(\bs C_B): \{\bs C_B \bs F=\bs 0\}=\{B_j \equiv 0 \; \forall j=1,\ldots,b\}$ for the main effect $B$.
Finally, the null hypothesis for the interaction is given by $H_0^F(\bs C_{AB}): \{\bs C_{AB} \bs F= \bs 0\}=\{(AB)_{ij} \equiv 0 \; \forall i=1,\ldots,a \text{ and } j=1,\ldots,b\}$.

\section{Testing procedures}\label{sec:testprod}

In this work, different testing procedures in ANOVA models are compared. First, classical ANOVA testing procedures with assumptions like normality or homoscedasticity are expounded and afterwards, more enhanced semi- and nonparametric techniques are illustrated. 

The classical one-way ANOVA $F$-test rejects the null hypothesis $H_0^\mu: \mu_1 = \ldots = \mu_d$ in the model~\eqref{equ:oneway} if
\begin{linenomath}
\[\frac{\frac{1}{d-1} \sum\limits_{i=1}^d n_i \left(\overline{X}_{i\cdot}-\overline{X}_{\cdot\cdot}\right)^2}{\frac{1}{N-d}\sum\limits_{i=1}^d \sum\limits_{k=1}^{n_i}\left(X_{ik}-\overline{X}_{\cdot\cdot}\right)^2} > F_{d-1, N-d, 1-\alpha},\]
\end{linenomath}
where $\overline{X}_{i\cdot}=\frac{1}{n_i}\sum_{j=1}^{n_i} X_{ij},$ $\overline{X}_{\cdot\cdot}=\frac{1}{N}\sum_{i=1}^d \sum_{j=1}^{n_i} X_{ij}$ and $F_{d-1, N-d, 1-\alpha}$ denotes the ($1-\alpha$)-quantile of an $F$-distribution with $d-1$ and $N-d$
degrees of freedom. It is an exact level-$\alpha$ testing procedure if 
$\varepsilon_{ik}\stackrel{\text{i.i.d.}}{\sim} N(0,\sigma^2)$ holds in \eqref{equ:oneway} for $\sigma^2 > 0$.
Otherwise, it is known to result in liberal or conservative conclusions \cite<see, e.g.,>{box1954some, deshon1996alternative, vallejo2010analysis, rosopa2013managing}.

To circumvent this issue, the application of  semi- or nonparametric approaches is usually recommended. 

\subsection{Semiparametric Methods}\label{sec:semimod}

Two standard semiparametric approaches were introduced in the work of \citeA{brunner1997box}.  These two test statistics are adequate inference procedures for testing null hypotheses formulated in terms of the means 
$H_0^\mu: \bs C \bm{\mu}=\bs 0$ in general factorial designs where $\bs C$ is an adequate contrast matrix (e.g. $\bs C=\bs P_d$ for testing $H_0^\mu$ in Equation~\eqref{equ:H0mu} in a one-way design or $\bs C \in \{\bs C_A, \bs C_B,
\bs C_{AB}\}$ in a two-way layout). Note, that in most cases it is possible to formulate other hypothesis than \eqref{equ:H0mu} using different contrast matrices $\bs C$. The first one is the so-called Wald-type statistic (WTS). Assuming a general heteroscedastic model \eqref{equ:oneway} and summarizing 
the original observations in a mean vector 
$\overline{\bs X}_\cdot = (\overline{X}_{1\cdot}, \ldots, \overline{X}_{d\cdot})'$, the WTS is given by
\begin{linenomath}
\[Q_N(\bs C) = N \overline{\bs X}_\cdot'  \bs C' \left(\bs C \widehat{\bs S}_N \bs C'\right)^+ \bs C \overline{\bs X}_\cdot. \]
\end{linenomath}
Here, a heteroscedastic covariance matrix estimator is given by $\widehat{\bs S}_N=N \cdot \diag\left(\frac{\widehat{\sigma}_1^2}{n_1},\ldots,\frac{\widehat{\sigma}_d^2}{n_d}\right)$ and $(\bs C)^+$ denotes the Moore-Penrose inverse of a matrix 
$\bs C$. Under the
null, the WTS is asymptotically $\chi^2$-distributed with $\rank(\bs C)$ degrees of freedom and, thus, the (classical) WTS procedure rejects $H_0^\mu$ if $Q_N(\bs C)$ is larger than $\chi^2_{\rank(\bs C), 1-\alpha}$, the $(1-\alpha)$-quantile of a 
$\chi^2$-distribution with $\rank(\bs C)$ degrees of freedom.

The second test statistic considered in the paper by \citeA{brunner1997box} is the so-called ANOVA-type statistic (ATS), which is based on a \citeA{box1954some}-type approximation. Assuming that $\tr(\bs D_M  \bs S_N)\neq 0$, the test statistic is given by
\begin{linenomath}
\[F_N(\bs M)=\frac{N}{\tr(\bs D_M \widehat{\bs S}_N)} \overline{\bs X}'_\cdot \bs M \overline{\bs X}_\cdot,\]
\end{linenomath}
where $\bs D_M=\diag(\bs M)$ with projection matrix $\bs M=\bs C' \left(\bs C \bs C'\right)^{-1} \bs C$. Since $F_N(\bs M)$ is asymptotically no pivot,  \citeA{brunner1997box} proposed to approximate its distribution by a central 
$F(\widehat{f}, \widehat{f}_0)$-distribution, where the degrees of freedom
are calculated by
\begin{linenomath}
\begin{align}\label{equ:ATSdof}
 \widehat{f}=\frac{\left[\tr(\bs D_M \widehat{\bs S}_N)\right]^2}{\tr(\bs M \widehat{\bs S}_N\bs M \widehat{\bs S}_N)} \;  \text{ and } \;  \widehat{f}_0=\frac{\left[\tr(\bs D_M \widehat{\bs S}_N)\right]^2}{\tr(\bs D_M^2 \widehat{\bs S}_N^2 \bs \Lambda)},
\end{align}
\end{linenomath}
with $\bm{\Lambda}=\diag\left((n_1-1)^{-1}, \ldots, (n_d-1)^{-1}\right)$. An advantage of the ATS (in contrast to the WTS) is that it can also handle singular covariance matrices $\widehat{\bs S}_N$. However, it is generally only an approximate
testing procedure. Another difference between these two test statistics is the unequal behavior in simulations. The ATS tends to conservative behavior while the WTS often exhibits very liberal results for small to moderate sample sizes \cite{vallejo2010analysis, pauly2015asymptotic}.   

To improve the small sample behavior of the WTS, \citeA{pauly2015asymptotic} proposed a permutation version of the WTS (WTPS). It is based on critical values obtained from the permutation distribution of $Q_N(\bs C)$, randomly permuting the pooled sample. 
All these procedures are applicable to arbitrary factorial designs  with fixed factors and implemented in the \textsc{R}-package \texttt{GFD} \cite<see>{GFD}.

\subsection{Nonparametric Methods}

In this section, the focus lies on nonparametric methods, namely the Kruskal-Wallis test, the VDW test and the rank-based versions of the ATS and the WTS known from the previous section
(rATS and rWTS). The first two tests assume equal variances among the groups, whereas the latter two can also deal with heteroscedasticity.

The Kruskal-Wallis test is widely known as the nonparametric equivalent of the ANOVA $F$-test for the analysis of one-way layouts of the form \eqref{equ:nonparamod} assuming a shift model (i.e. $F_i(x)=F_1(x-\mu_i)$ for $i=2,\ldots,d$). As many 
nonparametric approaches, it is based on ranks. Therefore, let  $R_{ik}$ be the midrank of $X_{ik}$ among all $N$ 
observations and $\overline{R}_{i\cdot}=\frac{1}{n_i}\sum_{k=1}^{n_i}R_{ik}$ the mean within group~$i$. 

Now, the test statistic of the Kruskal-Wallis test is given by
\begin{linenomath}
\begin{align} \label{equ:KW}
K_N &= \sum_{i=1}^d n_i \cdot \left(\overline{R}_{i\cdot}-\tfrac{N+1}{2}\right)^2 / \widehat{\sigma}_{H_0}^2,
\end{align}
\end{linenomath}
where $\widehat{\sigma}_{H_0}^2 = \frac{1}{N^2(N-1)}\sum_{i=1}^d \sum_{k=1}^{n_i} \left(R_{ik}-\frac{N+1}{2}\right)^2$. When the data is continuous, the estimator of the variance reduces to $\widehat{\sigma}_{H_0}^2 = \tfrac{N(N+1)}{12}$. The precision of
the Kruskal-Wallis statistic depends on the number of ties in the data and the number of groups \cite<see>[p. 103]{brunner2000nonparametric}. However, since the data is exchangeable under $H_0^F$, the exact distribution of the Kruskal-Wallis statistic (see 
Equation~\ref{equ:KW}) can be achieved by using its permutation distribution. Both versions -- the exact and the asymptotic one -- are implemented in \textsc{R}. One option to call up the method is using the command \texttt{kruskal\_test} implemented in the
package \texttt{coin} \cite<see>{zeileis2008implementing}. The asymptotic version is accessed by default and by using the argument \texttt{distribution=approximate(B=10000)} the permutation-based version with 10,000 permutation runs is conducted. 

The second method, which was also developed for shift models, is the van der Waerden (VDW) test. As the Kruskal-Wallis test, the VDW test is only applicable in the one-sample layout and is based on an inverse normal transformation. 
Defining the scores $A_{ik}:=\Phi^{-1}\left(\frac{R_{ik}}{n+1}\right)$, where $\Phi$ denotes the
distribution function of the normal distribution, the test statistic of the VDW test is given by
\begin{linenomath}
\begin{equation}
 \label{equ:vdW} T=\frac{1}{s^2}\sum_{i=1}^d n_i \overline{A}_i^2,
\end{equation}
\end{linenomath}
where $\overline{A}_i^2=\frac{1}{n_i}\sum_{k=1}^{n_i}A_{ik}$ and $s^2=\frac{1}{N-1}\sum_{i=1}^d\sum_{k=1}^{n_i} A_{ik}^2$.
In recent simulation studies \cite<see, e.g.,>{sheskin2004handbook, luepsen2017comparison}, the VDW test leads to convincing results. Therefore, it is also included in the simulation study.

To introduce the nonparametric version of the WTS (rWTS) and the ATS (rATS), consider the nonparametric effect measure
\begin{linenomath}
\[p_i=\int GdF_i, \; i=1,\ldots,d,\] 
\end{linenomath}
where $G(x)=\frac{1}{d}\sum_{i=1}^d F_i(x)$ denotes the unweighted mean distribution function. Different to other nonparametric effects considered in the literature, the above effect is based on the unweighted mean distribution functions possess the 
advantage that they do not depend on sample sizes and are therefore model constants \cite<see also the discussion in>{brunner2018rank}.
Let the empirical normalized rank mean vector $\widehat{\bs p}=\left(\widehat{p}_1, \ldots, \widehat{p}_d\right)'$ be the estimator of $\bs p=\left(p_1, \ldots, p_d\right)'$. Its entries are given by 
$\widehat{p}_i=\frac{1}{d}\sum_{r=1}^d \frac{1}{n_r}\left(\overline{R}_{i\cdot}^{(i+r)}-\frac{n_i+1}{2}\right)$, 
where $\overline{R}_{i\cdot}^{(i+r)}$ denotes the midrank of $X_{ik}$ among all $n_r + n_i$ observations. 
Again using $\bs D_M=\diag(\bs M)$ and $\bs M=\bs C' \left(\bs C \bs C'\right)^{-1} \bs C$, the test statistic for the rATS is given by
\begin{linenomath}
\begin{equation}
 \label{equ:ATS}F_N(\bs M)= \frac{N}{\tr(\bs D_M \widehat{\bs V}_N)} \widehat{\bs p}' \bs M  \widehat{\bs p}
\end{equation}
\end{linenomath}
and can be reasonably approximated by a central $F(\widehat{f}, \widehat{f}_0)$ distribution \cite{brunner1997box}. Here, the degrees of freedom are calculated as in Equation~\eqref{equ:ATSdof} by replacing $\widehat{\bs S}_N$ with
$\widehat{\bs V}_N=N\cdot \diag\left(\frac{\widehat{s}_1^2}{n_1}, \ldots, \frac{\widehat{s}_d^2}{n_d}\right),$ where $\widehat{s}_i=\frac{1}{N^2\left(n_i-1\right)}\sum_{k=1}^{n_i}\left(R_{ik}-\overline{R}_{i\cdot}\right)^2$ is 
the empirical rank-based variance for group $i=1,\ldots,d$.


Moreover, the nonparametric version of the WTS is defined by
\begin{linenomath}
\begin{equation}
 \label{equ:WTS}Q_N(\bs C) = N \bs{\widehat{p}}' \bs C' \left(\bs C \widehat{\bs V}_N \bs C'\right)^+ \bs C \bs{\widehat{p}}.
\end{equation}
\end{linenomath}
Also in this case, \citeA{brunner1997box} show that the rWTS is asymptotically $\chi^2_{\rank(C)}$-distributed under $H_0^F$. 
The latter two rank-based inference procedures for testing $H_0^F$ are implemented in the \textsc{R}-package \texttt{rankFD} \cite<see>{rankFD}.

Additional to the rATS and the rWTS, another permutation approach is conducted. It is based on a rank-based permutation version of the rWTS \cite<rWTPS, see>[for details]{umlauft2017rank}.

\subsection{Dropped Testing Procedures}
We note that there exist plenty of other inference procedures for factorial designs which we did not include in our simulation study due to non-promising behavior in existing simulation results \cite{vallejo2010analysis, richter2003performing} or a lack
of implementation in the statistic software \textsc{R}. In particular the latter holds for the generalized Welch-James and the approximate degree of freedom test mentioned in the Introduction.
%
%

\subsection{Summary}

The following two tables summarize the procedures of the whole section. Table~\ref{tab:overview} gives an overview of the different testing procedures for the one-way design. This table shows which testing procedure 
is recommended in which situation.

\begin{table}[!ht]
\footnotesize
\centering
\caption{\label{tab:overview}Overview of the procedures for the one-way layout and situations for which they were developed/recommended.}
 \begin{tabular}{lccc}
 \toprule
   & \multicolumn{1}{c}{\textbf{parametric}} & \multicolumn{2}{c}{\textbf{semi- and nonparametric}} \\
   & & \multicolumn{1}{c}{\textbf{original data}} & \multicolumn{1}{c}{\textbf{rank-based}} \\
   \midrule
   \multirow{2}{*}{\textbf{homoscedasticity}} & ANOVA $F$-Test & & Kruskal-Wallis Test\\
   & & & VDW Test \\
   \midrule
   \multirow{3}{*}{\textbf{heteroscedasticity}} & \texttt{oneway.test()}& ATS & rATS \\
   &  & WTS & rWTS \\
   &  & WTPS & rWTPS \\
   \bottomrule
 \end{tabular}
\end{table}

Table~\ref{tab:hypos} gives a short overview of the different testing procedures and their corresponding null hypothesis.

\begin{table}[!ht]
\centering
\footnotesize
\caption{\label{tab:hypos}Null hypotheses for which the test procedures used in the simulations were originally developed.}
\makebox[\textwidth]{%
 \begin{tabular}{lcccc}
 \toprule
  & $H_0^\mu$ & $H_0^F$ \\
  \midrule
  ANOVA $F$-Test & $\times$ &\\
  ATS & $\times$ &  \\
  Kruskal-Wallis Test & &$\times$ \\
  \texttt{oneway.test()} & $\times$ & \\
  rATS & & $\times$ &\\
  rWTPS & & $\times$& \\
  rWTS & & $\times$ &\\
  VDW Test & &$\times$ &\\
  WTPS & $\times$  &\\
  WTS & $\times$  &\\
  \bottomrule
 \end{tabular}}
\end{table}

\section{Simulation Study} \label{sec:simu}
\subsection{General design}
The different approaches to analyze factorial models are compared within an extensive simulation study. Therefore, the maintenance of the nominal type-$I$ error rate under the null hypothesis is examined. The simulations are conducted with the help of 
the \textsc{R} computing environment, version 3.4.0 \cite<see>{r2017language} each with 10,000 simulations and -- if necessary -- 10,000 permutation runs. The standard ANOVA $F$-test is compared to 
six different semi- and nonparametric procedures (ATS, WTS, permutation version of the WTS (WTPS) and the corresponding rank-based versions: rATS, rWTS, rWTPS). Furthermore, 
in the one-way case 
the VDW test, the two versions of the Kruskal-Wallis test and the \texttt{oneway.test()} are considered as well. All simulations are conducted for a significance
level of $\alpha=5\%$ and $\alpha=0.5\%$. First, the results of the simulations regarding $\alpha=5\%$ are given and in Section~\ref{sec:005} the results for $\alpha=0.5\%$ are presented.

\subsubsection{One-way layout}

We consider a one-way layout with five independent groups, where the observations are simulated by a shift-scale model 
\begin{linenomath}
\begin{equation}\label{equ:shiftscale}X_{ik}=\mu_i + \sigma_i\cdot\varepsilon_{ik}, \; i=1,\ldots,5; \; k=1,\ldots,n_i.\end{equation} 
\end{linenomath}
All group-specific means are set to zero ($\mu_1=\mu_2=\ldots=\mu_5=0$) in order that the null hypothesis $H_0^\mu$ in Equation~\eqref{equ:H0mu} is satisfied. Moreover, the nonparametric hypothesis $H_0^F: F_1=\ldots=F_5$ is definitely satisfied if the scaling factors 
$\sigma_1 = \ldots = \sigma_5$ are all equal (homoscedastic designs). The random error terms follow different standardized distributions
\begin{linenomath}
\[\varepsilon_{ik}=\frac{\tilde{\varepsilon}_{ik}-\mathbb{E}\left(\tilde{\varepsilon}_{i1}\right)}{\sqrt{\var\left(\tilde{\varepsilon}_{i1}\right)}}, \; i=1,\ldots,5; \; k=1,\ldots,n_i,\] 
\end{linenomath}
where the random variables $\tilde{\varepsilon}_{ik}$ were generated from normal, exponential, $\chi^2$, logistic, gamma and Poisson distributions. This covers a diverse range of different symmetric, skewed, 
continuous and discrete distributions.

The different scenarios are grouped in three different settings. The first setting includes  all symmetric distributions, such as the normal and the logistic distribution (see Table~\ref{tab:setting1}). The second setting deals with skewed distributions.
In this simulation study, we compare three different skewed distributions (exponential, $\chi^2$ and $\Gamma$) with different parameters in the different scenarios. All scenarios of Setting~2 are summarized in Table~\ref{tab:setting2}. The third setting 
examines the behavior of the different methods regarding discrete data. Therefore, a Poisson distribution is used to generate the data. An overview of the scenarios of Setting~3 is given in Table~\ref{tab:setting3}.

\begin{table}[!ht]
\centering
\footnotesize
\caption{\label{tab:setting1}Different scenarios of the first setting in the one-way layout indicating the underlying distribution, the sample size vector, the scaling factors and a short interpretation of the nine different scenarios.}
 \begin{tabular}{cllll}
 \toprule
  scenario & distribution & sample size & scaling & meaning \\
  \midrule
  1 & $N(0,1)$ & $\bs n =(5,5,5,5,5)'+m$ & $\bm{\sigma}=(1,1,1,1,1)'$ & balanced-homoscedastic \\[0.5em]
  2 & $N(0,1)$ & $\bs n =(5,5,5,5,15)'+m$ & $\bm{\sigma}=(1,1,1,1,1)'$ & unbalanced-homoscedastic \\[0.5em]
  3 & $N(0,1)$ & $\bs n =(5,5,5,5,5)'+m$ & $\bm{\sigma}=(1,1.2,1.5,1.7,2)'$ & balanced-heteroscedastic \\[0.5em]
  \multirow{2}{*}{4} & \multirow{2}{*}{$N(0,1)$} & \multirow{2}{*}{$\bs n =(4,7,10,13,15)'+m$} & \multirow{2}{*}{$\bm{\sigma}=(1,1.2,1.5,1.7,2)'$} & unbalanced-heteroscedastic  \\
    & & & & (positive pairing)\\[0.5em]
  \multirow{2}{*}{5} & \multirow{2}{*}{$N(0,1)$} & \multirow{2}{*}{$\bs n =(4,7,10,13,15)'+m$} & \multirow{2}{*}{$\bm{\sigma}=(2,1.7,1.5,1.2,1)'$} & unbalanced-heteroscedastic \\
  & & & & (negative pairing) \\[0.5em]
  \multirow{2}{*}{6} & 90\% $N(0,1)$    & \multirow{2}{*}{$\bs n =(5,5,5,5,5)'+m$} & \multirow{2}{*}{$\bm{\sigma}=(1,1,1,1,1)'$} & balanced-homoscedastic \\
  & \& 10\% $N(10,1)$ & & & with 10\% outlier\\[0.5em]
   \multirow{2}{*}{7} & 80\% $N(0,1)$    & \multirow{2}{*}{$\bs n =(5,5,5,5,5)'+m$} & \multirow{2}{*}{$\bm{\sigma}=(1,1,1,1,1)'$} & balanced-homoscedastic \\
  & \& 20\% $N(10,1)$ & & & with 20\% outlier\\[0.5em]
  \multirow{2}{*}{8} & \multirow{2}{*}{$\Logis(0,1)$} & \multirow{2}{*}{$\bs n =(4,7,10,13,15)'+m$} & \multirow{2}{*}{$\bm{\sigma}=(1,1.2,1.5,1.7,2)'$} & unbalanced-heteroscedastic  \\
  & & & & (positive pairing) \\[0.5em]
    \multirow{2}{*}{9} & \multirow{2}{*}{$\Logis(0,1)$} & \multirow{2}{*}{$\bs n =(4,7,10,13,15)'+m$} & \multirow{2}{*}{$\bm{\sigma}=(2,1.7,1.5,1.2,1)'$} & unbalanced-heteroscedastic  \\
  & & & & (negative pairing) \\
  \bottomrule
 \end{tabular}
\end{table}

\begin{table}[!ht]
 \centering
\footnotesize
\caption{\label{tab:setting2}Different scenarios of the second setting in the one-way layout indicating the underlying distribution, the sample size vector and a short interpretation of the 16 different scenarios; the scaling factor $\sigma=(1,1,1,1,1)'$ is the same in all 
scenarios.}
 \begin{tabular}{cllll}
 \toprule
  scenario & distribution & sample size& meaning \\
    \midrule
      1 & $\exp(1)$ & $\bs n =(5,5,5,5,5)'+m$  & balanced-homoscedastic \\
  2 & $\exp(1)$ & $\bs n =(5,5,5,5,15)'+m$  & unbalanced-homoscedastic \\
  3 & 90\% $\exp(1)$ \& 10\% $\exp(\frac{1}{10})$   & $\bs n =(5,5,5,5,5)'+m$  & balanced-homoscedastic with 10\% outlier \\
  4 & 80\% $\exp(1)$ \& 20\% $\exp(\frac{1}{10})$   & $\bs n =(5,5,5,5,5)'+m$  & balanced-homoscedastic with 20\% outlier \\
  5 & $\chi^2_3$ & $\bs n =(5,5,5,5,5)'+m$  & balanced-homoscedastic \\
  6 & $\chi^2_3$ & $\bs n =(5,5,5,5,15)'+m$  & unbalanced-homoscedastic \\
  7 & $\chi^2_{10}$ & $\bs n =(5,5,5,5,5)'+m$  & balanced-homoscedastic \\
  8 & $\chi^2_{10}$ & $\bs n =(5,5,5,5,15)'+m$  & unbalanced-homoscedastic \\
  9 & $\Gamma(10,0.1)$ & $\bs n=(5,5,5,5,5)'+m$ & balanced-homoscedastic\\
  10 & $\Gamma(10,0.1)$ & $\bs n=(5,5,5,5,15)'+m$ & unbalanced-homoscedastic\\
  11 & 90\% $\Gamma(1,0.1)$ \& 10\% $\Gamma(10,0.1)$ & $\bs n=(5,5,5,5,5)'+m$ & balanced-homoscedastic with 10\% outlier \\
  12 & 80\% $\Gamma(1,0.1)$ \& 20\% $\Gamma(10,0.1)$ & $\bs n=(5,5,5,5,5)'+m$ & balanced-homoscedastic with 20\% outlier \\
  13 & $\Gamma(1,2)$ & $\bs n=(5,5,5,5,5)'+m$ & balanced-homoscedastic\\
  14 & $\Gamma(1,2)$ & $\bs n=(5,5,5,5,15)'+m$ & unbalanced-homoscedastic\\
  15 & 90\% $\Gamma(1,2)$ \& 10\% $\Gamma(10,2)$ & $\bs n=(5,5,5,5,5)'+m$ & balanced-homoscedastic with 10\% outlier \\
  16 & 80\% $\Gamma(1,2)$ \& 20\% $\Gamma(10,2)$ & $\bs n=(5,5,5,5,5)'+m$ & balanced-homoscedastic with 20\% outlier \\
 \bottomrule
 \end{tabular}
\end{table}

\begin{table}[!ht]
 \centering
\footnotesize
\caption{\label{tab:setting3}Different scenarios of the third setting in the one-way layout indicating the underlying distribution, the sample size vector and a short interpretation of the four different scenarios; the scaling factor $\sigma=(1,1,1,1,1)'$ is the same in all 
scenarios.}
 \begin{tabular}{clll}
 \toprule
  scenario & distribution & sample size & meaning \\
    \midrule
  1 & $\Poi(25)$ & $\bs n=(5,5,5,5,5)'+m$ & balanced-homoscedastic\\
  2 & $\Poi(25)$ & $\bs n=(5,5,5,5,15)'+m$ & unbalanced-homoscedastic\\
  3 & 90\% $\Poi(25)$ \& 10\% $\Poi(5)$ & $\bs n=(5,5,5,5,5)'+m$ & balanced-homoscedastic with 10\% outlier \\
  4 & 80\% $\Poi(25)$ \& 20\% $\Poi(5)$ & $\bs n=(5,5,5,5,5)'+m$ & balanced-homoscedastic with 20\% outlier \\

 \bottomrule
 \end{tabular}

\end{table}

To increase the sample size, a constant $m \in \{0,5,10,15,20,25\}$ was added to each component of the sample size vector $\bs n$. 
In the past, often difficulties with the maintenance of the nominal type-$I$ error rate in unbalanced-heteroscedastic designs (no matter if positive or negative pairing) were detected \cite<see, e.g.,>{vallejo2010robust, umlauft2017rank}. 
Thus, such settings are also included in the 
simulation. 
Moreover, the behavior of the different procedures when outliers are present should be examined. Therefore, different scenarios where 10\%, 15\% or 20\% outliers are present in the data are simulated. For example, in the sixth scenario of the first setting
10\% outliers are simulated while adding 10\% random variables of the overall sample size simulated by a normal distribution with mean 10 (instead of 0 as in the other 90\%) and variance one.

\subsubsection{Two-way layout}

For a crossed two-way layout, data is also generated by a shift-scale model $X_{ijk}=\mu_{ij}+\sigma_{ij}\cdot\varepsilon_{ijk}, \; i=1,\ldots,a, \; j=1,\ldots,b, \;k=1,\ldots,n_{ij}$ and as in the previous section, the error terms
follow a standardized normal, exponential, $\chi^2$, logistic, gamma or Poisson distribution. The simulations are restricted to $2 \times 5$-layouts with balanced sample size vectors $\bs n=(\bs n_3',\bs n_3')'$, where 
$\bs n_3=(n_{31}, \ldots, n_{35})'=(5,5,5,5,5)'$ as well as unbalanced sample size vectors $\bs n=(\bs n_1', \bs n_2')'$ with $\bs n_1=(n_{11},\ldots,n_{15})'=(4,4,4,4,4)'$ and $\bs n_2=(n_{21},\ldots,n_{25})'=(7,7,7,7,7)'$. To increase the sample size,
a constant $m \in \{0,5,10,15,20,25\}$ is added to each component of the sample size vector $\bs n$.

\begin{table}[!ht]
\centering
\footnotesize
\caption{\label{tab:setting1_twoway}Different scenarios of the first setting in the two-way layout indicating the underlying distribution, the sample size vector, the scaling factors and a short interpretation of the eight different scenarios.}
 \begin{tabular}{cllll}
 \toprule
  scenario & distribution & sample size & scaling & meaning \\
  \midrule
  1 & $N(0,1)$ & $\bs n =(\bs n_3',\bs n_3')'+m$ & $\bm{\sigma}=(1,\ldots,1,1,\ldots,1)'$ & balanced-homoscedastic \\[0.5em]
  2 & $N(0,1)$ & $\bs n =(\bs n_1',\bs n_2')'+m$ & $\bm{\sigma}=(1,\ldots,1,1,\ldots,1)'$ &unbalanced-homoscedastic \\[0.5em]
  3 & $N(0,i^2)$ & $\bs n =(\bs n_3',\bs n_3')'+m$ & $\bm{\sigma}=(1,\ldots,1, 2,\ldots,2)'$ &balanced-heteroscedastic \\[0.5em]
  \multirow{2}{*}{4} & \multirow{2}{*}{$N(0,i^2)$} & \multirow{2}{*}{$\bs n =(\bs n_1',\bs n_2')'+m$} &\multirow{2}{*}{$\bm{\sigma}=(1,\ldots,1, 2,\ldots,2)'$} & unbalanced-heteroscedastic  \\
     & & & &(positive pairing)\\[0.5em]
  \multirow{2}{*}{5} & \multirow{2}{*}{$N(0,i^2)$} & \multirow{2}{*}{$\bs n =(\bs n_2',\bs n_1')'+m$} &\multirow{2}{*}{$\bm{\sigma}=(2,\ldots,2, 1,\ldots,1)'$} & unbalanced-heteroscedastic \\
   & & && (negative pairing) \\[0.5em]
      \multirow{2}{*}{6} & 85\% $N(0,1)$ & \multirow{2}{*}{$\bs n =(\bs n_3',\bs n_3')'+m$} &  \multirow{2}{*}{$\bm{\sigma}=(1,\ldots,1,1,\ldots,1)'$} &balanced-homoscedastic  \\
   & \& 15\% $N(10,1)$ & & &with 15\% outlier \\[0.5em]
     \multirow{2}{*}{7} & \multirow{2}{*}{Logistic(0,$i^2$)} & \multirow{2}{*}{$\bs n =(\bs n_1',\bs n_2')'+m$} & \multirow{2}{*}{$\bm{\sigma}=(1,\ldots,1, 2,\ldots,2)'$} & unbalanced-heteroscedastic  \\
   & & & &(positive pairing) \\[0.5em]
    \multirow{2}{*}{8} & \multirow{2}{*}{Logistic(0,$i^2$)} & \multirow{2}{*}{$\bs n =(\bs n_2',\bs n_1')'+m$} & \multirow{2}{*}{$\bm{\sigma}=(2,\ldots,2, 1,\ldots,1)'$} & unbalanced-heteroscedastic  \\
   & & & &(negative pairing) \\
  \bottomrule
 \end{tabular}
\end{table}

\begin{table}[!ht]
 \centering
\footnotesize
\caption{\label{tab:setting2_twoway}Different scenarios of the second setting in the two-way layout indicating the underlying distribution, the sample size vector and a short interpretation of the 13 different scenarios; the scaling factor $\sigma=(1,\ldots,1)'$ is the same in all 
scenarios.}
 \begin{tabular}{clll}
 \toprule
  scenario & distribution & sample size & meaning \\
    \midrule
      1 & $\exp(1)$ & $\bs n =(\bs n_3',\bs n_3')'+m$  & balanced-homoscedastic \\
  2 & $\exp(1)$ & $\bs n =(\bs n_1',\bs n_2')'+m$ & unbalanced-homoscedastic \\
  3 & 85\% $\exp(1)$ \& 15\% $\exp(\frac{1}{10})$& $\bs n =(\bs n_3',\bs n_3')'+m$  & balanced-homoscedastic with 15\% outlier  \\
  4 & $\chi^2_3$ & $\bs n =(\bs n_3',\bs n_3')'+m$ & balanced-homoscedastic \\
  5 & $\chi^2_3$ & $\bs n =(\bs n_1',\bs n_2')'+m$  & unbalanced-homoscedastic \\
  6 & $\chi^2_{10}$ & $\bs n =(\bs n_3',\bs n_3')'+m$ & balanced-homoscedastic \\
  7 & $\chi^2_{10}$ & $\bs n =(\bs n_1',\bs n_2')'+m$  & unbalanced-homoscedastic \\
  8 & $\Gamma(10,0.1)$ & $\bs n=(\bs n_3', \bs n_3')'+m$ & balanced-homoscedastic\\
  9 & $\Gamma(10,0.1)$ & $\bs n=(\bs n_1',\bs n_2')'+m$ & unbalanced-homoscedastic\\
  10 & 85\% $\Gamma(1,0.1)$ \& 15\% $\Gamma(10,0.1)$ & $\bs n=(\bs n_3',\bs n_3')'+m$ & balanced-homoscedastic with 15\% outlier \\
  11 & $\Gamma(1,2)$ & $\bs n=(\bs n_3',\bs n_3')'+m$ & balanced-homoscedastic\\
  12 & $\Gamma(1,2)$ & $\bs n=(\bs n_1',\bs n_2')'+m$ & unbalanced-homoscedastic\\
  13 & 85\% $\Gamma(1,2)$ \& 15\% $\Gamma(10,2)$ & $\bs n=(\bs n_3',\bs n_3')'+m$ & balanced-homoscedastic with 15\% outlier \\
 \bottomrule
 \end{tabular}
\end{table}

\begin{table}[!ht]
 \centering
\footnotesize
\caption{\label{tab:setting3_twoway}Different scenarios of the third setting in the two-way layout indicating the underlying distribution, the sample size vector and a short interpretation of the three different scenarios; the scaling factor $\sigma=(1,\ldots,1)'$ is the same in all 
scenarios.}
 \begin{tabular}{clll}
 \toprule
  scenario & distribution & sample size & meaning \\
    \midrule
  1 & $\Poi(25)$ & $\bs n=(\bs n_3,\bs n_3)+m$ & balanced-homoscedastic\\
  2 & $\Poi(25)$ & $\bs n=(\bs n_1,\bs n_2)+m$ & unbalanced-homoscedastic\\
  3 & 85\% $\Poi(25)$ \& 15\% $\Poi(5)$ & $\bs n=(\bs n_3,\bs n_3)+m$ & balanced-homoscedastic with 15\% outlier \\
 \bottomrule
 \end{tabular}
\end{table}

As above, the first setting summarizes the scenarios regarding all symmetric distributions considered in this simulation study, namely the normal and the logistic distribution. The second setting deals with skewed distributions and covers three 
distributions (exponential, $\chi^2$ and $\Gamma$) and thirteen different scenarios including balanced and unbalanced designs as well as scenarios with and without outliers in the generated data. The third setting deals with discrete data, therefore, the data 
is again generated by a Poisson distribution. The variation of the different settings are summarized in Tables~\ref{tab:setting1_twoway}~-~\ref{tab:setting3_twoway}, where Table~\ref{tab:setting1_twoway} describes all scenarios of Setting 1 (symmetric 
distributions), Table~\ref{tab:setting2_twoway} of Setting 2 (skewed distributions) and Table~\ref{tab:setting3_twoway} of Setting 3 (discrete distributions). The notation within the tables of this section stays the same as in the tables of the one-way layout.

\subsection{Results}
\subsubsection{One-way layout}

\begin{figure}[!ht]
 \centering
 \includegraphics[width=\textwidth ]{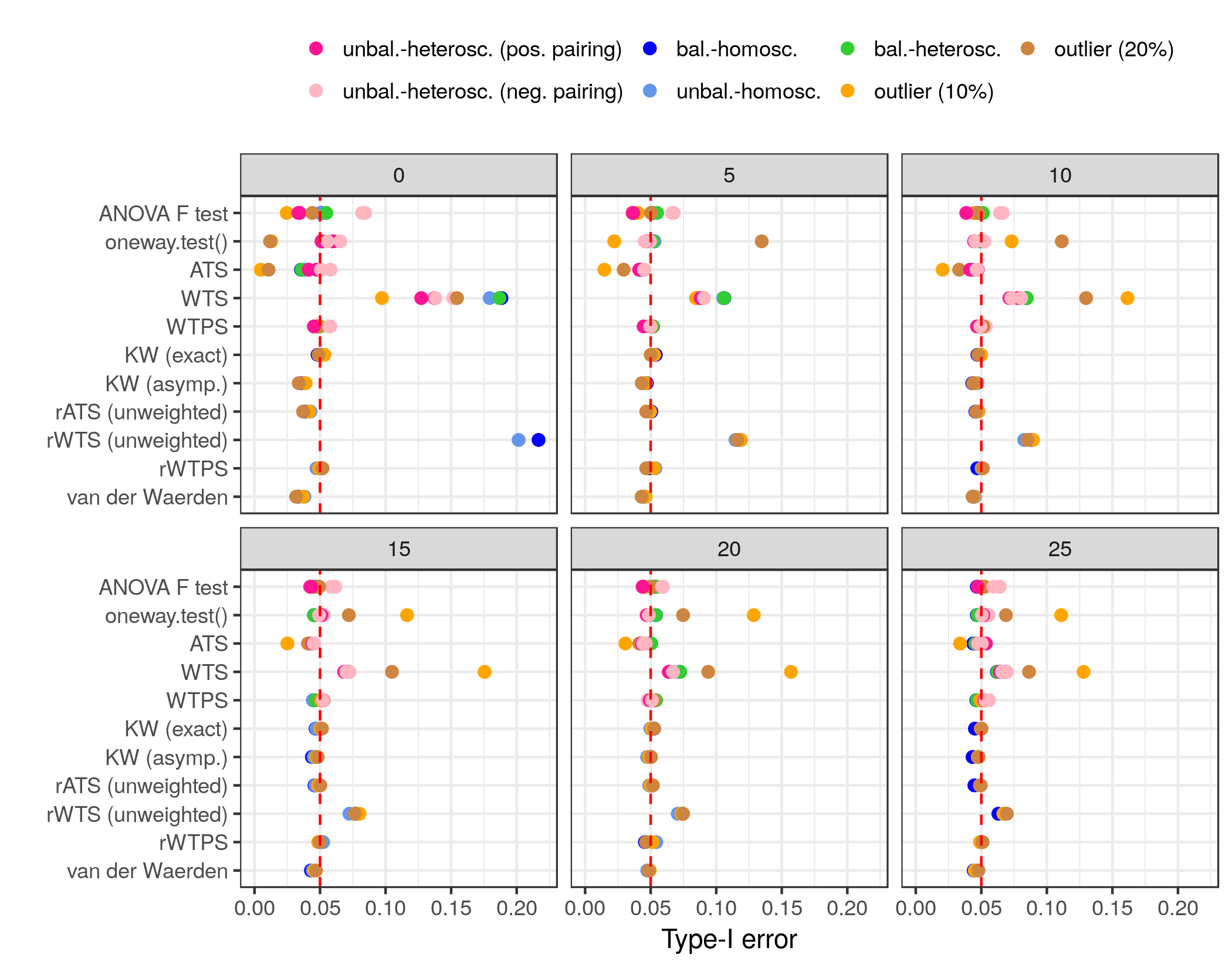}
 \caption{\label{fig:set1} \bf Type-$I$ error rates at 5\% for all considered scenarios in the first setting regarding the one-way layout.}
\end{figure}

\begin{figure}[!ht]
 \centering
\includegraphics[width=\textwidth ]{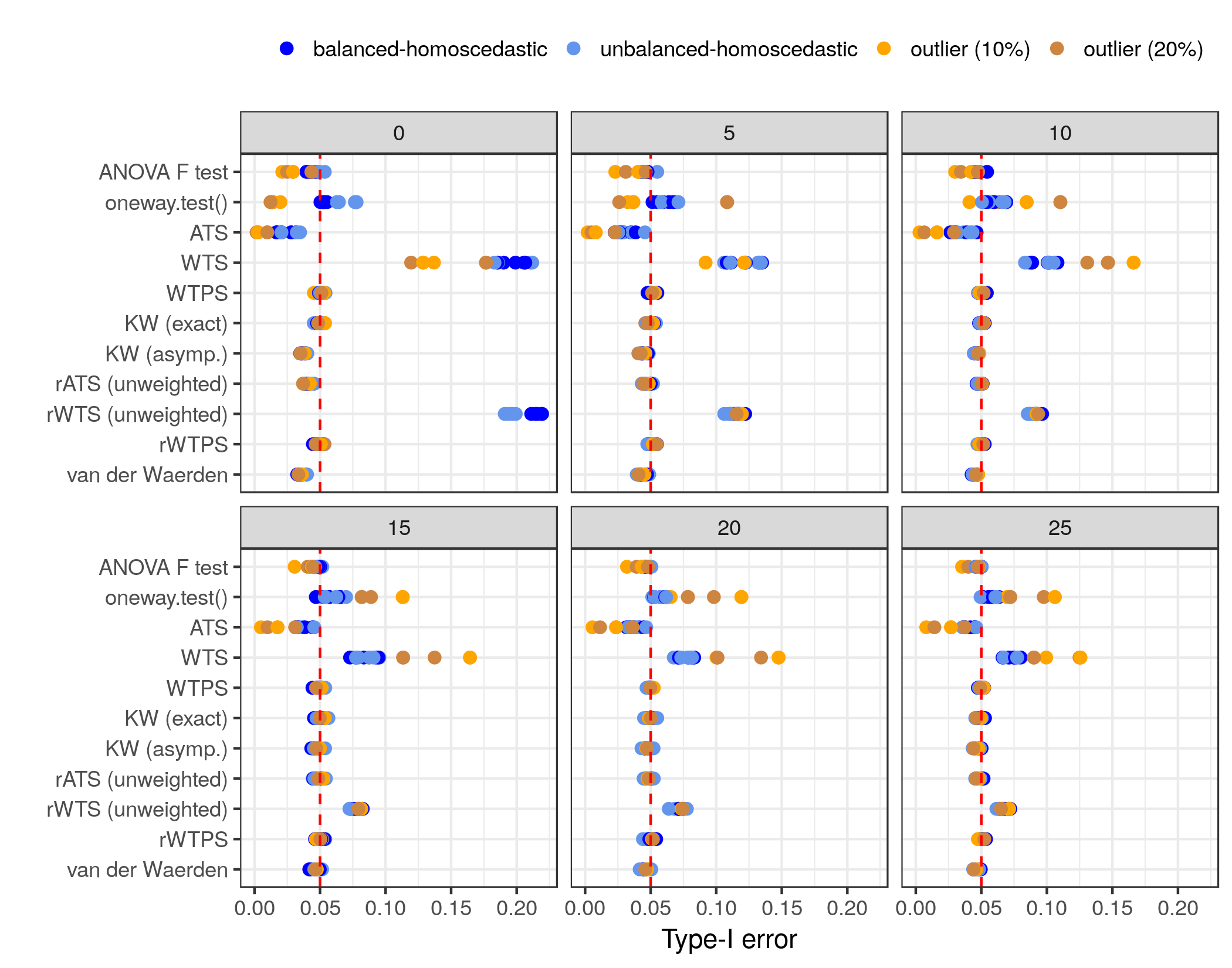}
 \caption{\label{fig:set2} \bf Type-$I$ error rates at 5\% for all considered scenarios in the second setting regarding the one-way layout.}
\end{figure}

\begin{figure}[!ht]
 \centering
 \includegraphics[width=\textwidth ]{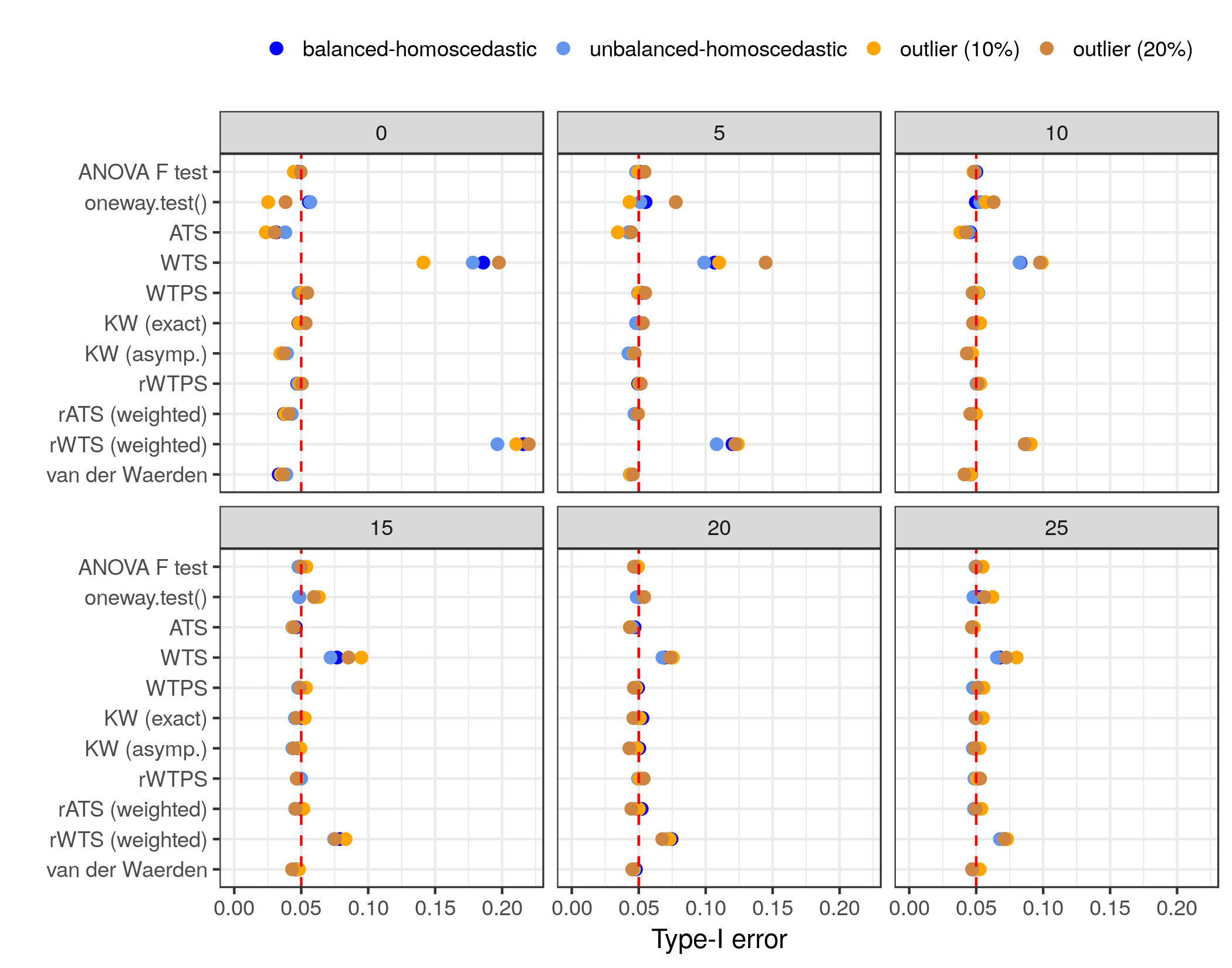}
 \caption{\label{fig:set3} \bf Type-$I$ error rates at 5\% for all considered scenarios in the third setting regarding the one-way layout.}
\end{figure}

Here, the results of the type-$I$ error rates for the one-way layout are presented. In particular, the results for the first setting (symmetric distributions) are given in Figure~\ref{fig:set1}, for the second setting (skewed distributions) in 
Figure~\ref{fig:set2} and for the third setting (discrete distributions) in Figure~\ref{fig:set3}. In these figures, the type-$I$ errors of the different scenarios are displayed by colored dots. Each type of scenario has its individual color to detect those ones 
which either tend to conservative or liberal behavior. 
Thus, the balanced-homoscedastic designs are colored in blue, the unbalanced-homoscedastic designs in light blue, balanced-heteroscedastic designs in green, the positive paired scenarios of the unbalanced-heteroscedastic design in pink, the negative 
pairing in light pink and the outlier settings are given in yellowish colours, whereas scenarios including 10\% and 15\% outliers are presented in orange and scenarios where 20\% outliers are present are colored in brown. Each
row in the figures summarize one different testing procedure and the six different panels show the results for the constant $m \in \{0,5,\ldots,25\}$, which is used to increase each component of the sample size vector $\bs n$. The vertical red line displays the underlying 
statistical significance level $\alpha$. Thus, all dots located on the left side of this line indicate a conservative behavior of the test, whereas dots on the right side might imply the liberality of the presented test statistic. To make the analysis and
interpretation of the figures more simple, the parametric and semiparametric test are located in the upper part of each panel and the nonparametric testing procedures are presented in the lower part.

As already mentioned in the section about the statistical model and the corresponding hypotheses, we have to distinguish between the two different hypothesis and have to ensure that both hypotheses hold true in the different scenarios since otherwise, we
do not gain reliable testing result. 
For the one-way layout, it is easy to see that the null hypothesis formulated in terms of means $H_0^\mu$ holds true in all scenarios since the shift parameter $\mu_i$ in the shift-scale design (Equation~\ref{equ:shiftscale}) was set to zero among all 
different groups. In case of hypotheses formulated in terms of distribution functions and heteroscedasticity, this specific null hypothesis $H_0^F$ does not hold true. This is an (at least interpretative) disadvantage of the rank procedures for the
null hypothesis $H_0^F$ which one should be aware of. An alternative to the nonparametric effects introduced in this work is given by more descriptive effect sizes \cite<see>{brunner2018rank}, which are not presented here, since we focused on the 
classical testing procedures. Therefore, the corresponding results are omitted in the following.

The summarized results for Setting 1 in the one-way layout (see Figure~\ref{fig:set1}) are as follows: 
The Wald-type statistic (WTS) shows very liberal results in all scenarios and especially for small to moderate sample sizes. But even for large sample sizes and e.g.~scenarios including outliers the type-$I$ error rate of the WTS is still very 
conservative. In contrast, the permutation-based version of the WTS (WTPS) controls the $\alpha$-level very well for all scenarios. The VDW test is conservative for small to moderate sample sizes and controls the type-$I$ error quite
good for larger sample sizes. The unweighted version of the rank-based WTS (rWTS) shows a liberal behavior for all simulated sample sizes and scenarios, whereas again the permuted version (rWTPS) reveals very accurate results for all scenarios and sample sizes. 
Quite as good as the rWTPS are the results of the unweighted version of the rank-based ANOVA-type statistic (rATS), only for small sample sizes the behaviors are not as accurate as for the rWTPS. The \texttt{oneway.test()} controls the type-$I$ error 
very good for most scenarios. Just dealing with scenarios where outliers are present lead to conservative as well as liberal test decisions. The two versions -- asymptotic and exact -- of the Kruskal-Wallis test (KW) show good results.
Generally, the exact version of the test works better than the asymptotic version. The classical ATS is very conservative even for larger sample sizes and for scenarios where
outliers are present. The ANOVA $F$-test which was given as a state-of-the-art test shows good results for most cases, but for scenarios where the assumptions of the ANOVA $F$-test are violated, inflated
type-$I$ errors show up. 

Figure~\ref{fig:set2} summarizes the results for Setting 2 (skewed distributions) in the one-way layout. Compared to Figure~\ref{fig:set1} the conclusions of the results are nearly the same. Therefore, only testing procedures with conspicuous behavior are 
mentioned in the following.
Again the WTS and the rWTS show very liberal results for small to moderate sample sizes. The behavior of the rank-based WTS is better for large sample sizes, whereas the WTS is not adequate even for large sample sizes. The \texttt{oneway.test()} has 
some difficulties in controlling the type-$I$ error rate and shows conservative as well as liberal results. As in Setting~1, the ATS is conservative for nearly all scenarios and sample sizes. The other testing procedures do not show any remarkable 
behavior.

The interpretation of Figure~\ref{fig:set3} is less sophisticatedly since there are only four modeled scenarios. Again, the WTS has a very liberal behavior, whereas the permutation version of the WTS controls the type-$I$ error accurately. The VDW test 
shows its  difficulties for small sample sizes, where it tends to conservative behavior.  Again, the rank-based Wald-type statistic is very liberal and the permutation version (WTPS) outperforms the rank-based WTS (rWTS) and shows very good results. The rank-based ATS 
tends to conservative results for very small sample sizes, but otherwise, it provides accurate results. The \texttt{oneway.test()} has some difficulties when outliers are present in the data. In such scenarios, it sometimes leads to conservative as well 
as liberal results. But for large sample sizes, the \texttt{oneway.test()} can handle this issue. Both versions of the Kruskal-Wallis test statistic lead to very good results; only for small sample sizes, the asymptotic test is somewhat conservative. The 
standard ATS is conservative for small to moderate sample sizes but controls the type-$I$ error rate for large sample sizes. The ANOVA $F$-test works well for all four scenarios and sample sizes.

\subsubsection{Two-way layout}
For the two-way layout, only the results for the interaction hypothesis are presented. Main effects can be investigated as in the one-way layout. Also in the two-way layout, we have to distinguish between scenarios when the hypotheses $H_0^\mu$ and 
$H_0^F$ hold true. As in the one-way layout, the interaction null hyptheses formulated in terms of the mean hold true in all simulated scenarios, whereas the interaction null hypotheses formulated in terms of distribution functions are still not fulfilled in 
scenarios where heteroscedasticity is present. Again, the user should keep in mind that the (interaction) null hypotheses formulated in terms of distribution functions have a different interpretation than the null hypotheses formulated in terms of means.

\begin{figure}[!ht]
 \centering
 \includegraphics[width=\textwidth ]{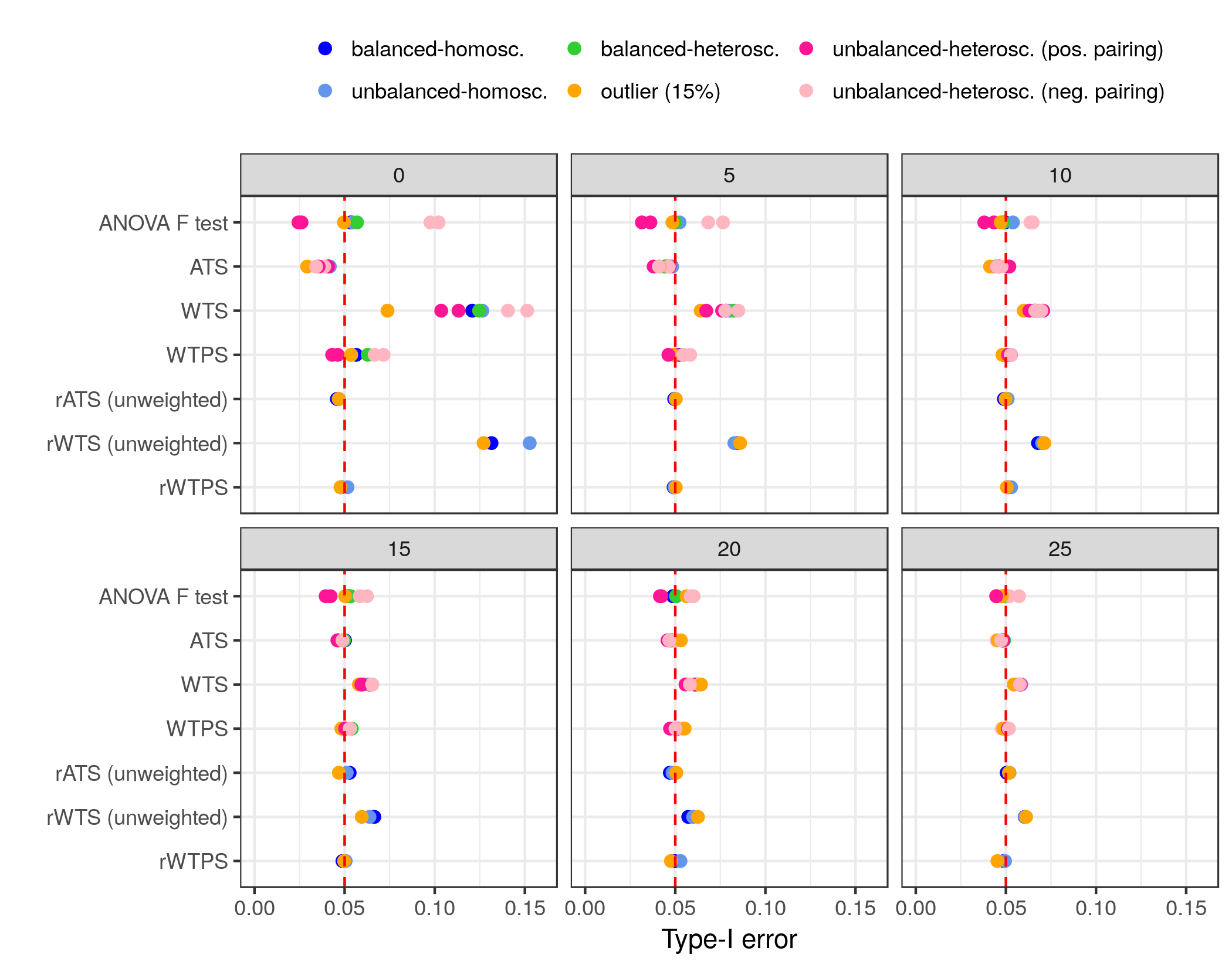}
 \caption{\label{fig:set1_two} \bf Type-$I$ error rates at 5\% for all considered scenarios in the first setting regarding the two-way layout.}
\end{figure}

\begin{figure}[!ht]
 \centering
 \includegraphics[width=\textwidth ]{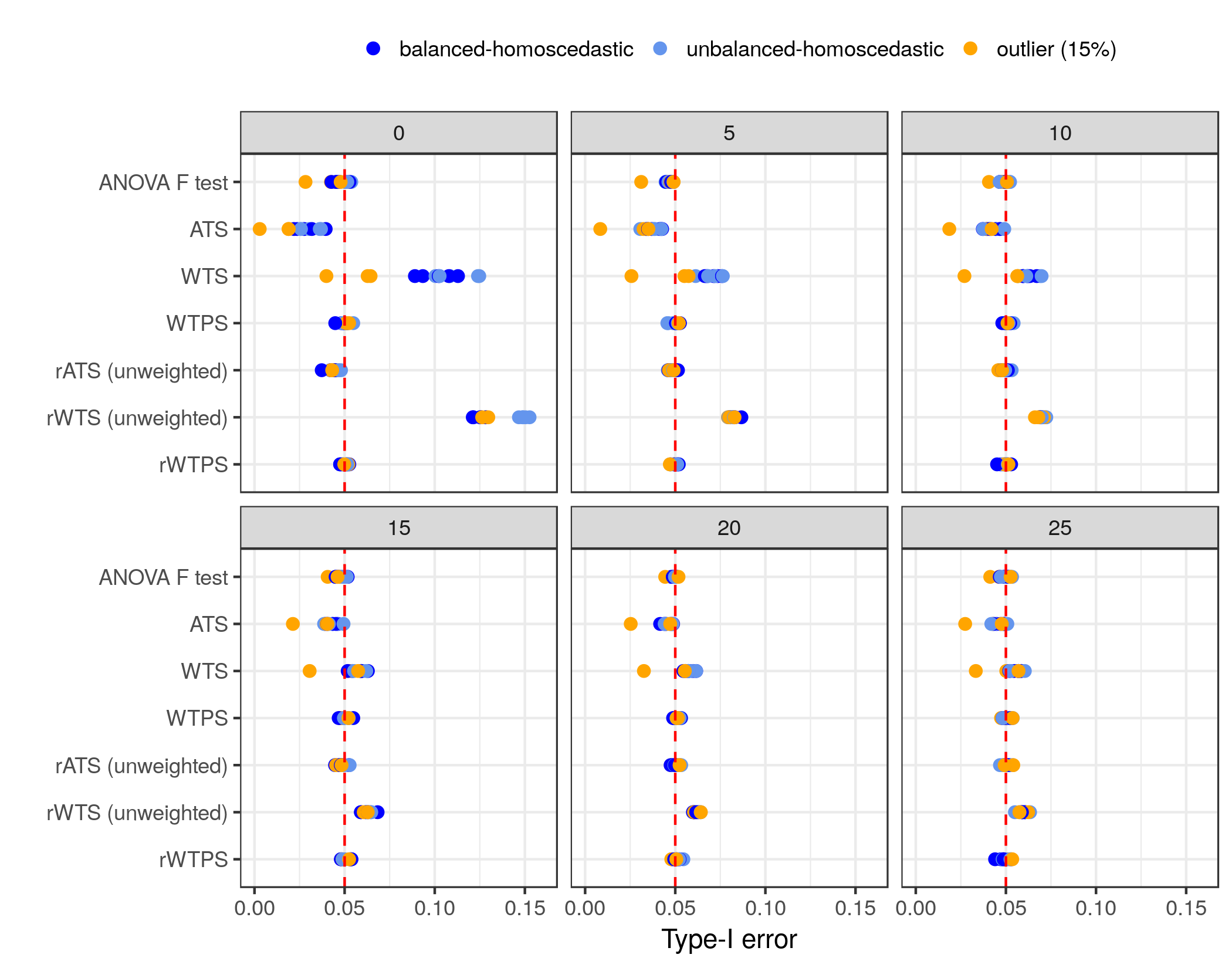}
 \caption{\label{fig:set2_two} \bf Type-$I$ error rates at 5\% for all considered scenarios in the second setting regarding the two-way layout.}
\end{figure}

\begin{figure}[!ht]
 \centering
 \includegraphics[width=\textwidth ]{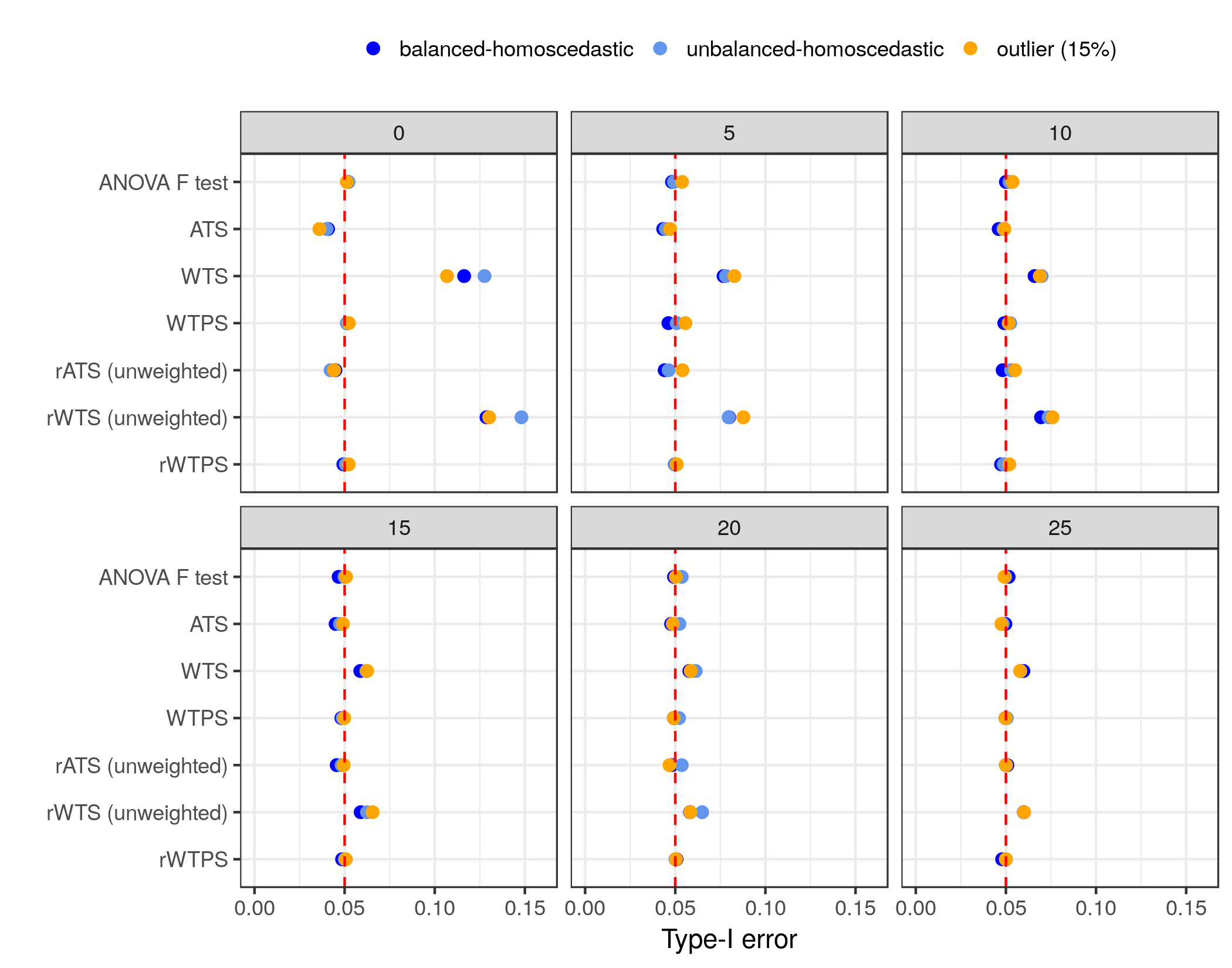}
 \caption{\label{fig:set3_two} \bf Type-$I$ error rates at 5\% for all considered scenarios in the third setting regarding the two-way layout.}
\end{figure}

Figure~\ref{fig:set1_two} summarizes the results of the first setting (symmetric distributions), 
Figure~\ref{fig:set2_two} of the second setting (skewed distributions) and Figure~\ref{fig:set3_two} of the third setting (discrete distributions).

The results of Setting 1 in the two-way layout are presented in Figure~\ref{fig:set1_two} and are described below. If the sample size decreases, the results of the classical WTS becomes more liberal. For small sample sizes, the permutation-version of the WTS shows conservativeness
for the scenarios with positive pairing and liberality for scenarios with negative pairing as well as for the balanced-heteroscedastic design. For moderate to large sample sizes, the results of the WTPS are convincing. 
The rank-based WTS in its unweighted version tends to very liberal behavior, whereas the permuted version of the rank-based WTS (rWTPS) works very good. The rank-based 
ATS controls the type-$I$ error rate very good. The classical ATS has some difficulties for small sample sizes. In such cases, the results are a little conservative. For small sample sizes, the ANOVA $F$-test does not 
control the type-$I$ error very accurate but for moderate to large sample sizes the type-$I$ error control is very reasonable.

Regarding the second setting (skewed distributions), the results as given in Figure~\ref{fig:set2_two} are as follows: The behavior of the WTS is liberal in most cases, but for larger sample sizes the results become better and thus, the type-$I$ error 
control of the WTS for larger sample sizes is quite good. Only for one unbalanced-homoscedastic design, the WTS shows a conservative result for all sample sizes. Nevertheless, very accurate results are obtained by the permutation version of the WTS (WTPS). Again, the rank-based WTS (rWTS) shows liberal results, whereas the rank-based Wald-type permutation 
statistic (rWTPS) controls the type-$I$ error rate very accurately as well as the rank-based ATS does. The ATS is conservative for small to moderate sample sizes and even for one unbalanced-homoscedastic design, it is very conservative among all sample
sizes. The results of the ANOVA $F$-test are reasonable, only for small sample sizes, one scenario shows a conservative result.

As above, the interpretation of Figure~\ref{fig:set3_two} -- and therefore Setting 3 (discrete data) -- is more simple, since only three scenarios are visualized. The WTS tends to very liberal results for small to moderate sample sizes and once more the 
permutation version of the WTS (WTPS) shows very good results in controlling the type-$I$ error. The rank-based WTS is very liberal, whereas the permuted version of the rank-based WTS controls the type-$I$ error level very accurately. As in the one-way 
layout, the rATS shows a very good $\alpha$-level control in all scenarios and sample sizes. The ATS works well, only for small sample sizes it shows a slight conservativeness. Furthermore, the classical way to analyze factorial models (ANOVA $F$-test) shows good results.

\subsubsection{Simulation results for $\alpha= $ 0.5\%}\label{sec:005}
In this section, all results for the simulations with a significance level of 0.5\% are presented. In the Appendix, we like to compare the deviation from the preannounced significance level for 
$\alpha=0.5\%$ and $\alpha=5\%$ as well.

\begin{figure}[!ht]
 \centering
 \includegraphics[width=\textwidth]{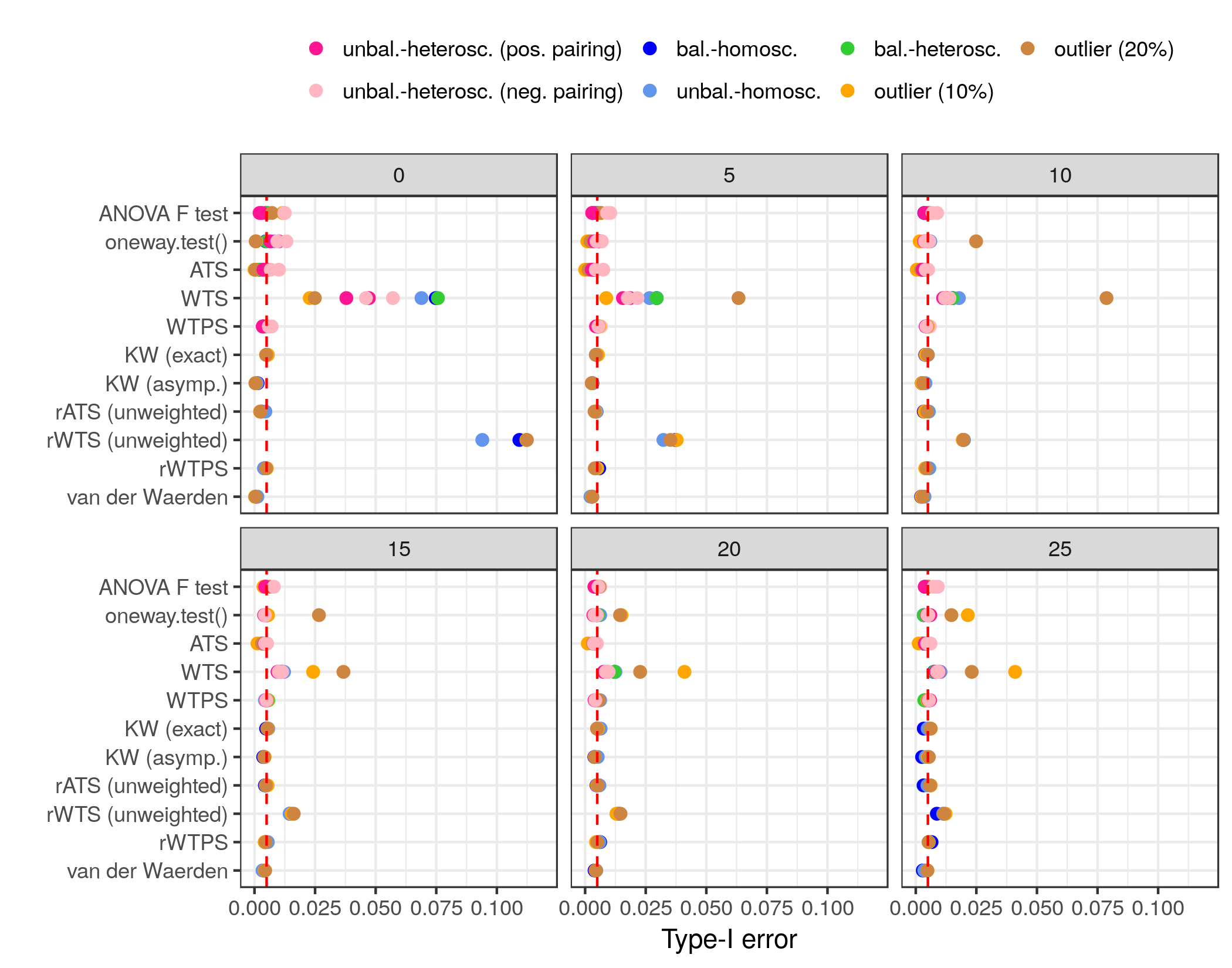}
 \caption{\label{fig:set1_005} \bf Type-$I$ error rates at 0.5\% for all considered scenarios in the first setting regarding the one-way layout.}
\end{figure}

\begin{figure}[!ht]
 \centering
 \includegraphics[width=\textwidth]{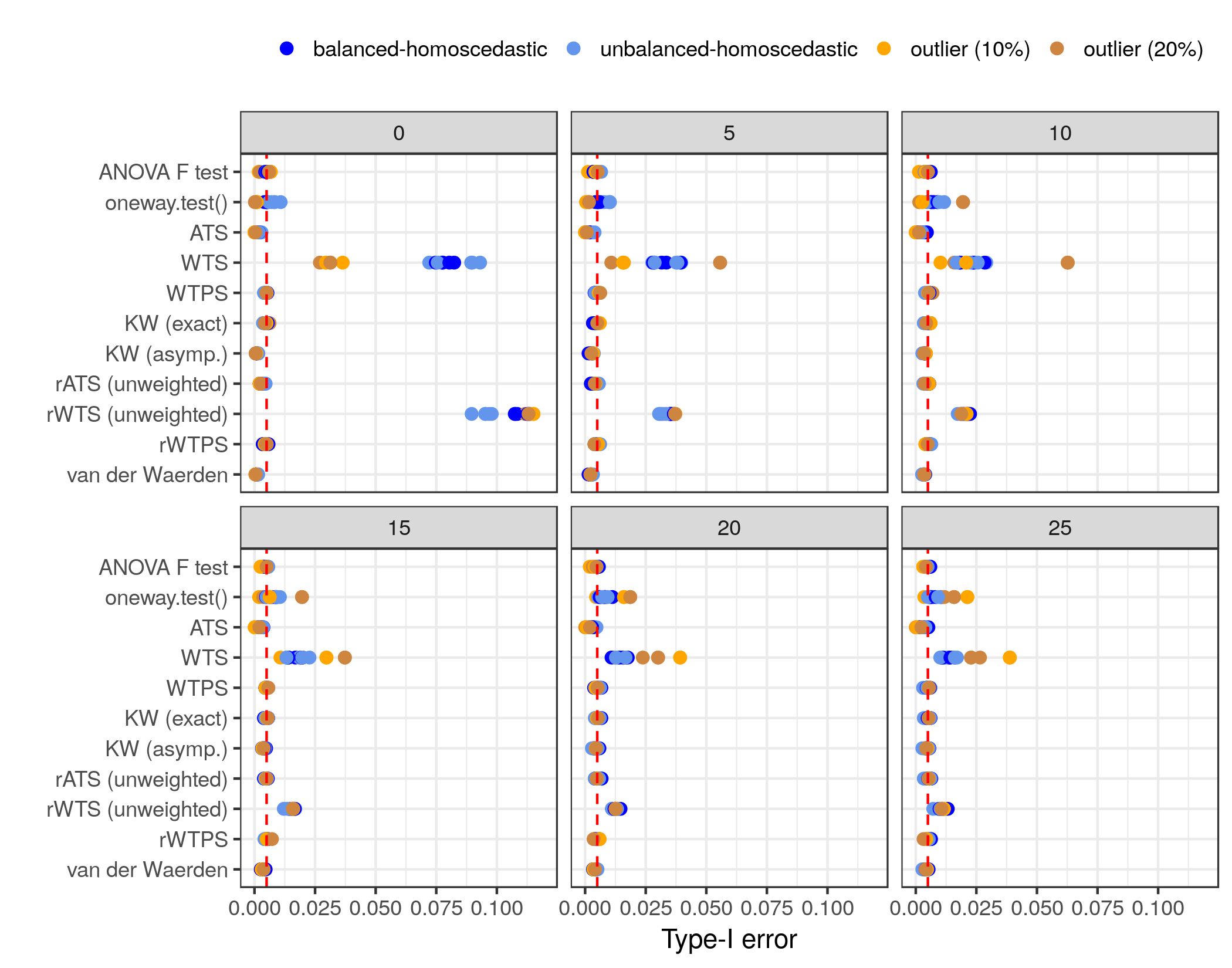}
 \caption{\label{fig:set2_005} \bf Type-$I$ error rates at 0.5\% for all considered scenarios in the second setting regarding the one-way layout.}
\end{figure}

\begin{figure}[!ht]
 \centering
 \includegraphics[width=\textwidth]{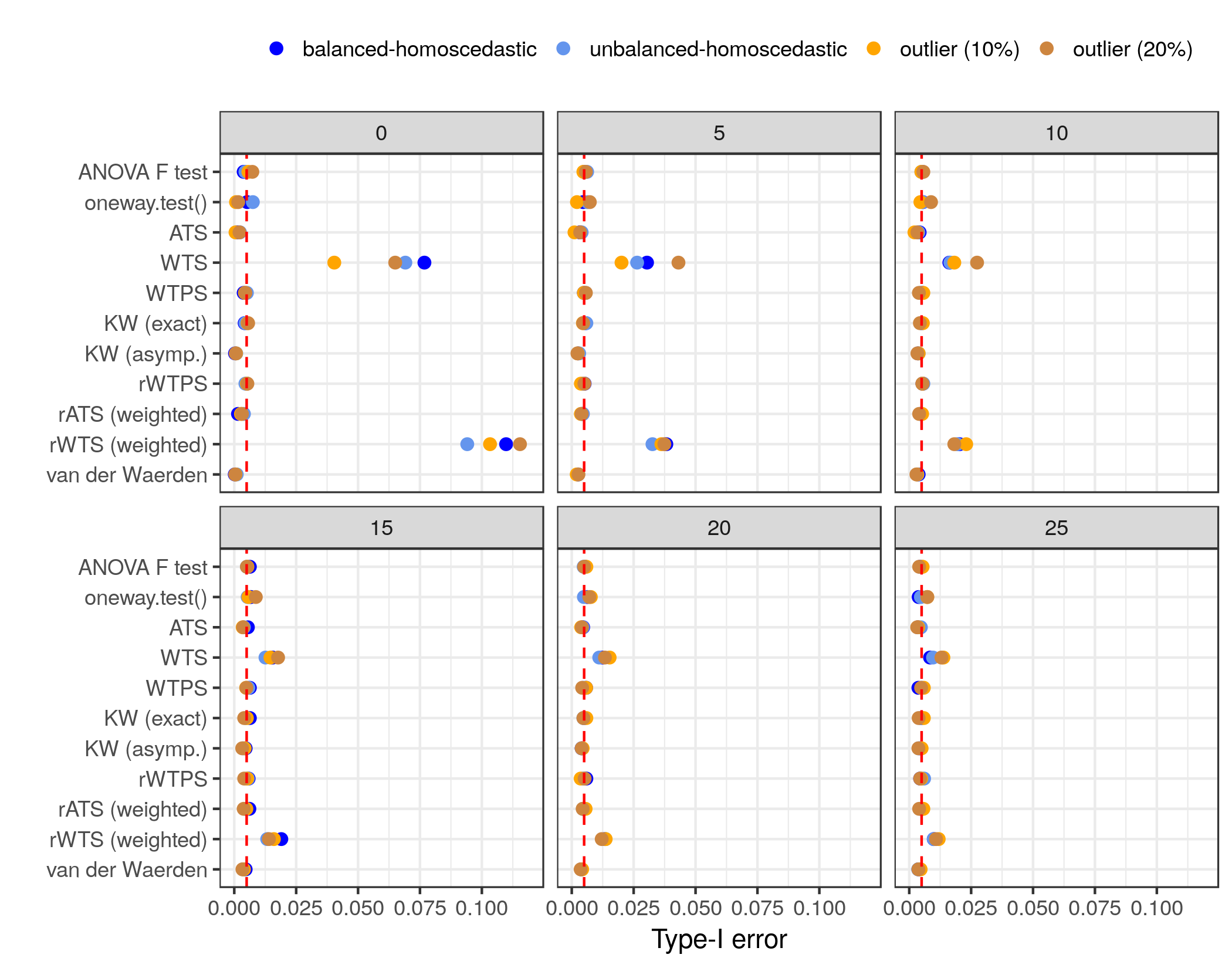}
 \caption{\label{fig:set3_005} \bf Type-$I$ error rates at 0.5\% for all considered scenarios in the third setting regarding the one-way layout.}
\end{figure}

\begin{figure}[!ht]
 \centering
 \includegraphics[width=\textwidth]{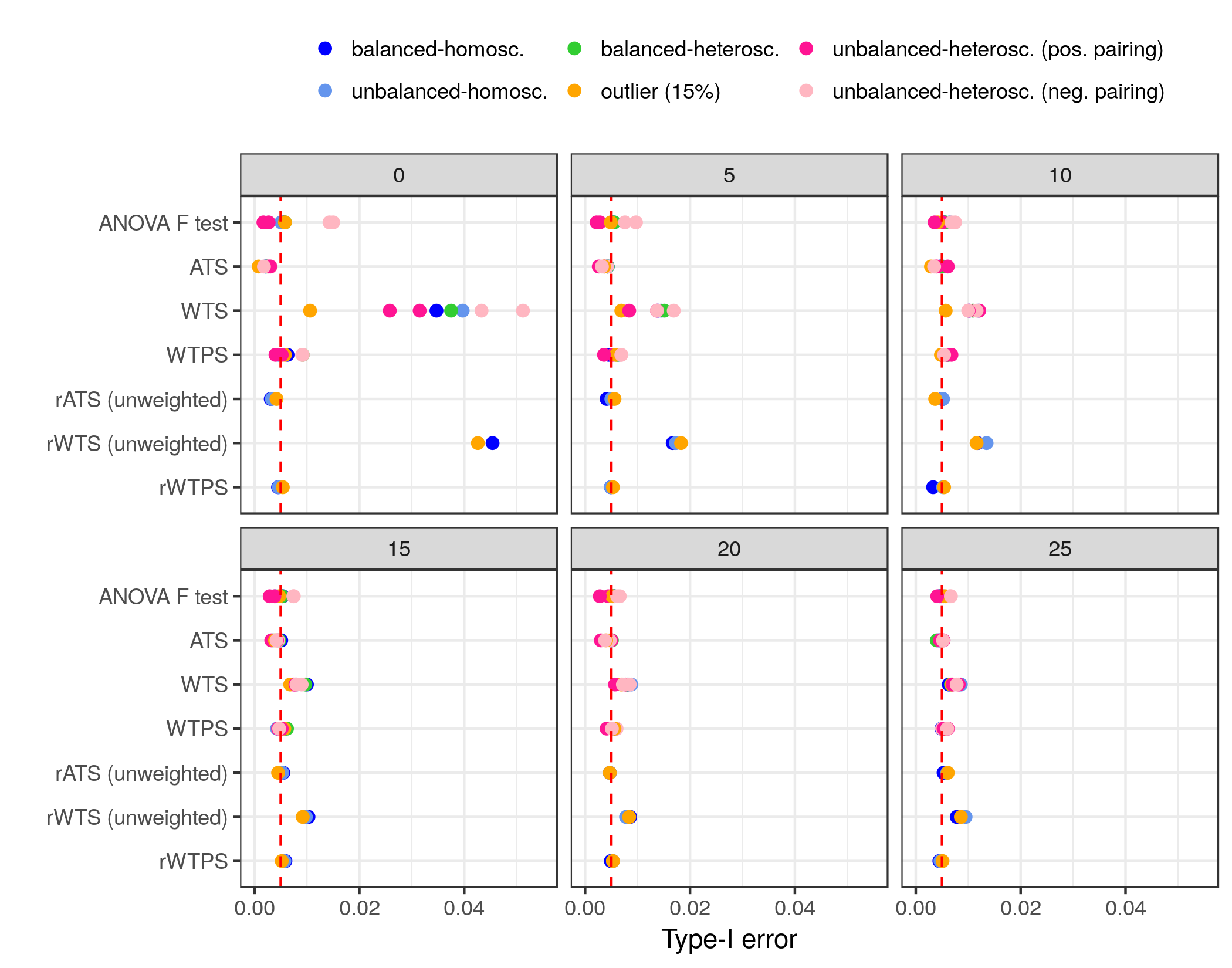}
 \caption{\label{fig:set1_two_005} \bf Type-$I$ error rates at 0.5\% for all considered scenarios in the first setting regarding the two-way layout.}
\end{figure}

\begin{figure}[!ht]
 \centering
 \includegraphics[width=\textwidth]{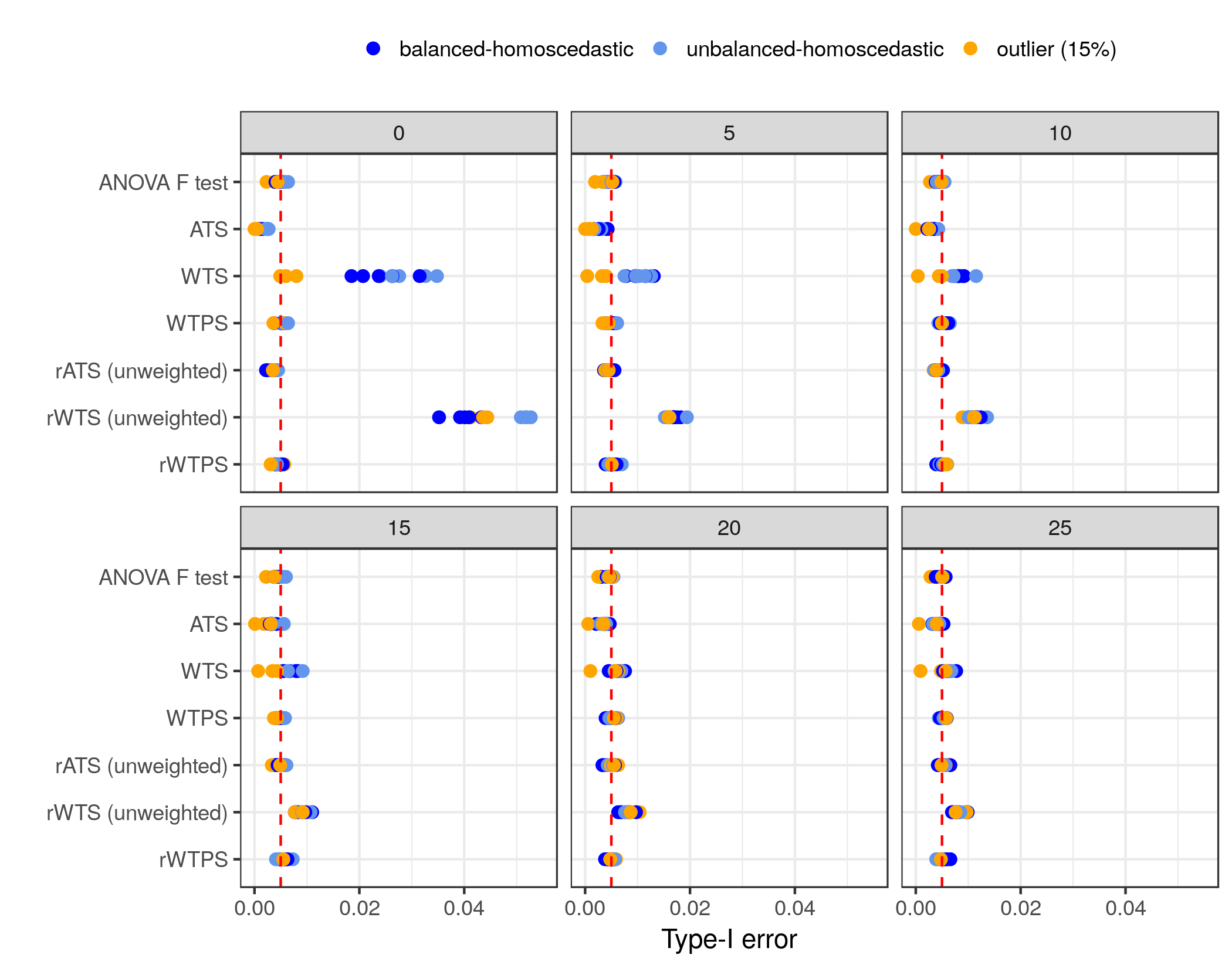}
 \caption{\label{fig:set2_two_005} \bf Type-$I$ error rates at 0.5\% for all considered scenarios in the second setting regarding the two-way layout.}
\end{figure}

\begin{figure}[!ht]
 \centering
 \includegraphics[width=\textwidth]{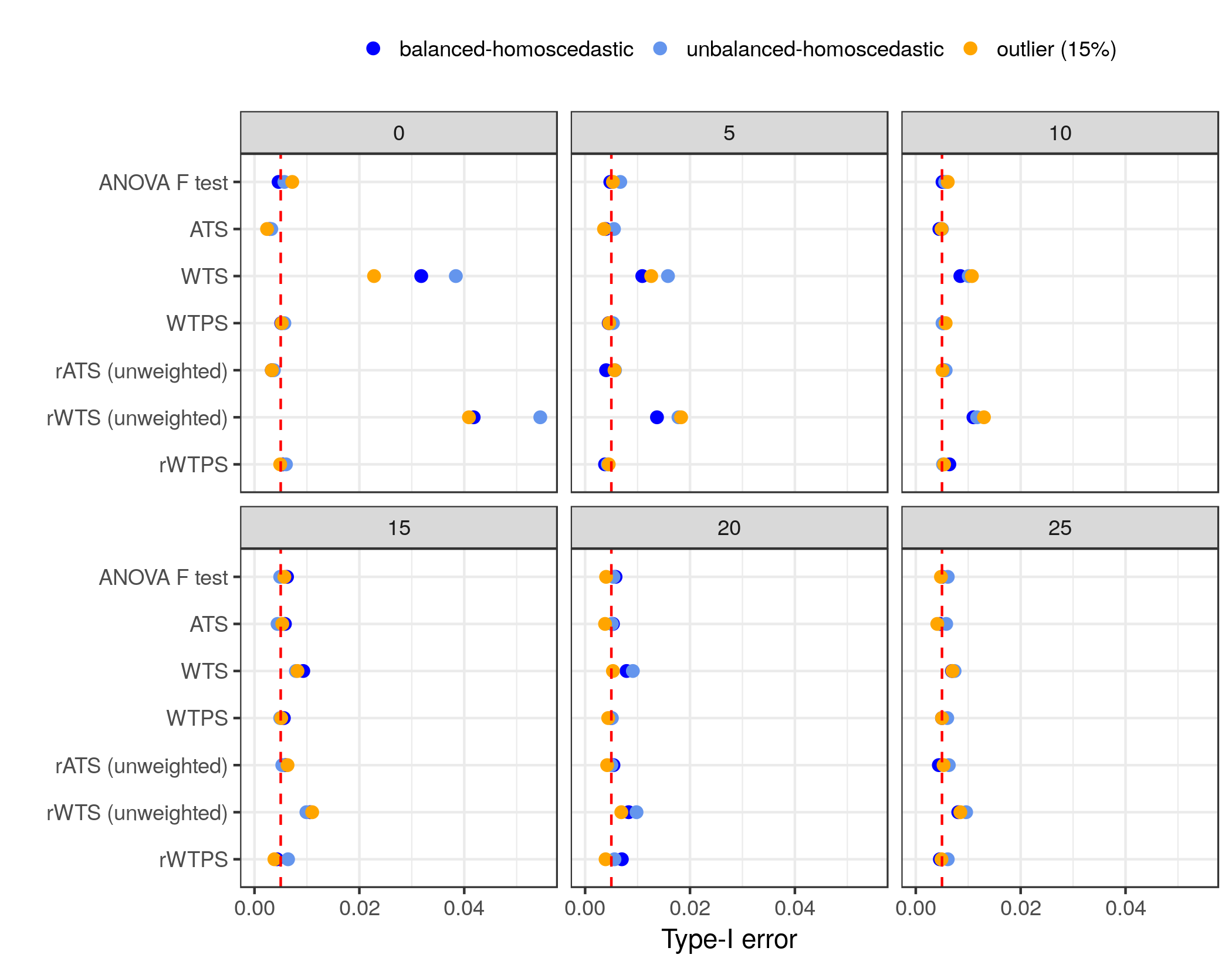}
 \caption{\label{fig:set3_two_005} \bf Type-$I$ error rates at 0.5\% for all considered scenarios in the third setting regarding the two-way layout.}
\end{figure}

Figures~\ref{fig:set1_005}-\ref{fig:set3_005} show the results for the one-way layout, whereas Figure~\ref{fig:set1_005} visualizes the results of the simulations regarding Setting 1 (symmetric distributions), Figure~\ref{fig:set2_005} for Setting 2
(skewed distributions) and Figure~\ref{fig:set3_005} 
for Setting 3 (discrete data). Similar to the one-way layout, the results of the two-way layout are given in Figures~\ref{fig:set1_two_005}-\ref{fig:set3_two_005}. Figure~\ref{fig:set1_two_005} summarizes the results of Setting 1, Figure~\ref{fig:set2_two_005} of 
Setting 2 and Figure~\ref{fig:set3_two_005} of Setting 3.
In the following, the different figures are shortly summarized. Since the underlying significance level of 0.5\% is very small it is hard to find conservative test decision. 

In Figure~\ref{fig:set1_005} the results of the first setting regarding the one-way layout and a type-$I$ error rate of 0.5\% are visualized. The results of most testing procedures are quite good. Therefore, only testing procedures which show difficulties
in controlling the type-$I$ error are described below. Again, the WTS shows liberal results for small to moderate sample sizes and also for large sample sizes in case of settings where outliers are present. Also the unweighted version of the rank-based
WTS is liberal for all sample sizes but becomes better if sample sizes are rising. The \texttt{oneway.test()} shows some difficulties in controlling the type-$I$ error when outliers are present and also for unbalanced-heteroscedastic designs.

Figure~\ref{fig:set2_005} deals with the one-way layout and the second setting (skewed distributions). The WTS again is liberal for all sample sizes and all considered scenarios. Also the rank-based WTS (rWTS) shows liberal results for small to moderate 
sample sizes. The \texttt{oneway.test()} shows scenarios with conservative results and other scenarios where the test leads to liberal results. For small sample sizes, the asymptotic version of the KW test and the ATS show conservative test decisions for 
nearly all scenarios.

The interpretation of Figure~\ref{fig:set3_005} (discrete data in the one-way layout) is again less complicated due to the fact that only four different scenarios were modeled. Again, we have to deal with the WTS and the rWTS because they tend to be 
liberal for all sample sizes. The VDW test leads to conservative results for small to moderate sample sizes as well as the asymptotic version of the Kruskal-Wallis test and the ATS. The \texttt{oneway.test()} shows difficulties in controlling the type-$I$
errors in scenarios where outliers are present. In such cases, the test reveals conservative as well as liberal results.

The next three figures deal with the results of the two-way layout. In particular, Figure~\ref{fig:set1_two_005} (Setting 1, symmetric distributions) show up that all testing procedures work well for large sample sizes. Only in case of small sample sizes,
the WTS and the rank-based versions of the WTS show some difficulties in controlling the type-$I$ errors.  The ANOVA $F$-test shows liberal as well as conservative test decisions. For negative paired unbalanced-heteroscedastic designs it is liberal 
for small to moderate 
sample sizes and for positive paired unbalanced-heteroscedastic designs it shows a conservative behavior for small to moderate sample sizes. The results of the ATS are conservative for small sample sizes.

The results of Setting 2 (skewed distributions) in the two-way layout are summarized in Figure~\ref{fig:set2_two_005}. Again, the WTS shows very liberal results, but for unbalanced-homoscedastic designs, the test tends to conservative results for all 
sample sizes. The rank-based WTS does not show this phenomenon and is again very liberal for small to moderate sample sizes. In contrast to this, the ATS tends to conservative results for all scenarios while dealing with small to moderate sample
sizes.

Figure~\ref{fig:set3_two_005} summarizes the results for Setting 3 (discrete distributions) in the two-way layout. Again, the WTS and the rank-based version of the WTS show liberal results for small to moderate sample sizes. For small sample sizes, the ATS
tends to conservative results. 



\section{Discussion and Conclusion} \label{sec:con}

In this work, various different testing procedures to analyze factorial models were introduced and finally compared during an extensive simulation study. Additionally, we focus on the behavior of the procedures while estimating very small quantiles,
since a large group of researchers proposed to redefine the significance level ($\alpha=0.5\%$) to improve the reproducibility \cite<see>{benjamin2017redefine}.

In the following, the discussion is devided into two parts -- the interpretation and discussion for parametric and semiparametric procedures is given first and afterwards, the nonparametric are discussed. A reason for
that devision is the underlying null hypothesis. In the previous sections, the theoretical and interpretational differences were already discussed. 

Regarding the parametric and semiparametric approaches, the Wald-type permutation-based test statistic (WTPS) shows the best results, since it keeps the preassigned significance level very accurate for most sample sizes and scenarios. For symmetric 
distribution, the ANOVA-type test statistic (ATS) controls the type-$I$ error rates, whereas in case of skewed distributions the results tend to be very conservative. For larger sample sizes, the ANOVA $F$-test shows good results if no outliers and no 
positive or negative pairing is present. 

In case of the considered nonparametric testing procedures, the best performance in the simulation study was shown by the rank-based permutation procedure (rWTPS) and for the one-way layout also the exact version of the Kruskal-Wallis 
test shows very accurate results in controlling the type-$I$ errors. The rank-based ANOVA-type test statistic leads to very accurate results for moderate to large sample sizes and is therefore recommended in such situations due to the fast computation 
time. Nevertheless, in case of very small sample sizes the permutation approaches, especially the rank-based versions, are the best choice. 

Also in case of a significance level of $\alpha=0.5\%$, the WTPS (in the semiparametric case), the rWTPS (in the nonparametric case) and the rank-based ATS show the best results. Moreover, for the one-way design, the VDW and
the Kruskal-Wallis testing procedure are also reasonable approaches. But with regard to the comparisons of the test results for $\alpha=5\%$ and $\alpha=0.5\%$, most procedures show some difficulties in controlling the type-$I$ error rates in case 
of small quantiles (here: $\alpha=0.5\%$).

To sum up, the nonparamtric testing procedures, especially the rank-based permutation Wald-type test statistic and the rank-based ANOVA-type test statistic shows the best results in our extensive simulation study. But as a limitation of the simulations
should be mentioned that all different scenarios reflect scenarios where the data is exchangeable and therefore, the corresponding nonparametric testing procedures are exact level-$\alpha$ tests. Thus, the simulated scenarios are more advantageous for
the nonparametric procedures. Addtionally, the user should be aware of the fact, that the null hypthesis formulated in terms of the mean and the hypothesis formulated in terms of distribution functions are not equivalent. Therefore, settings where
$H_0^\mu$ holds true, whereas $H_0^F$ does not and vice versa exist.

\section*{Acknowledgements}

The author would like to thank Markus Pauly for helpful discussions. 
The work of the author was supported by the German Research Foundation project DFG-PA 2409/3-1.

\nolinenumbers
\bibliographystyle{apacite}
\bibliography{biblio}

\newpage
\singlespacing
\appendix

\section{Comparisons between the deviation of the prerequired significance levels of 5\% and 0.5\%}

\doublespacing

This section shows a comparison of the results obtained in the simulations with a significance level of 5\% and with a significance level of 0.5\%. In the following figures (Figures \ref{fig:set1_abw}-\ref{fig:set3_abw_two}), the deviation (in percent)
of both prerequired levels are summarized. The red dots represent a level of 5\%, whereas the black dots summarize the results for a significance level of 0.5\%. To sum up, all six figures show that the simulation results regarding the larger 
significance level (5\%) reveal a better control of the type-$I$ error rates, whereas the black dots representing a level of 0.5\% highlight a very bad behavior in controlling the type-$I$ errors in some scenarios. 

\begin{figure}[!ht]
 \centering
 \includegraphics[width=\textwidth]{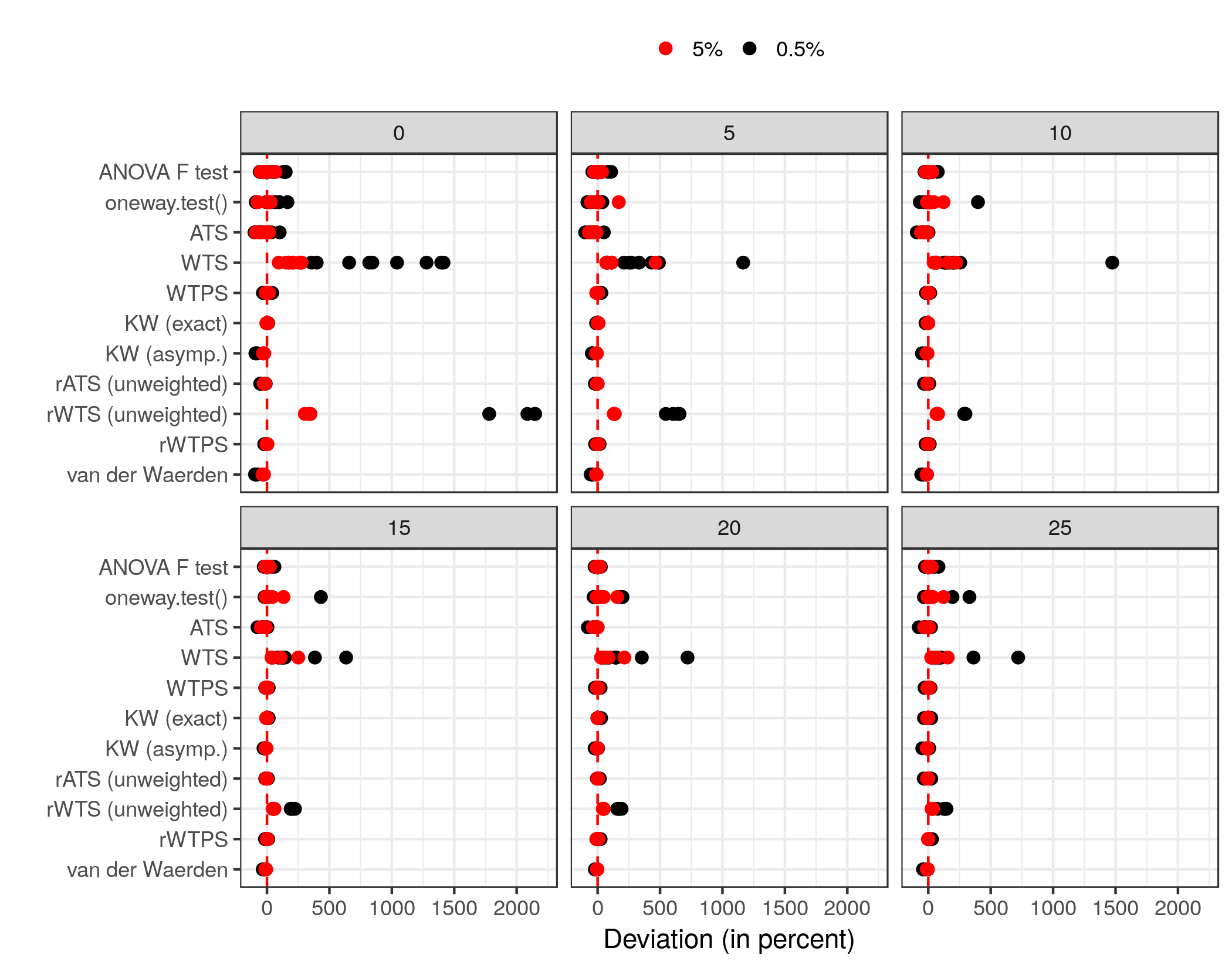}
 \caption{\label{fig:set1_abw} \bf Deviation (in percent) of the prerequired significance levels 5\% (red dots) and 0.5\% (black dots) in the first setting regarding the one-way layout.}
\end{figure}
\newpage
\begin{figure}[!ht]
 \centering
 \includegraphics[width=\textwidth]{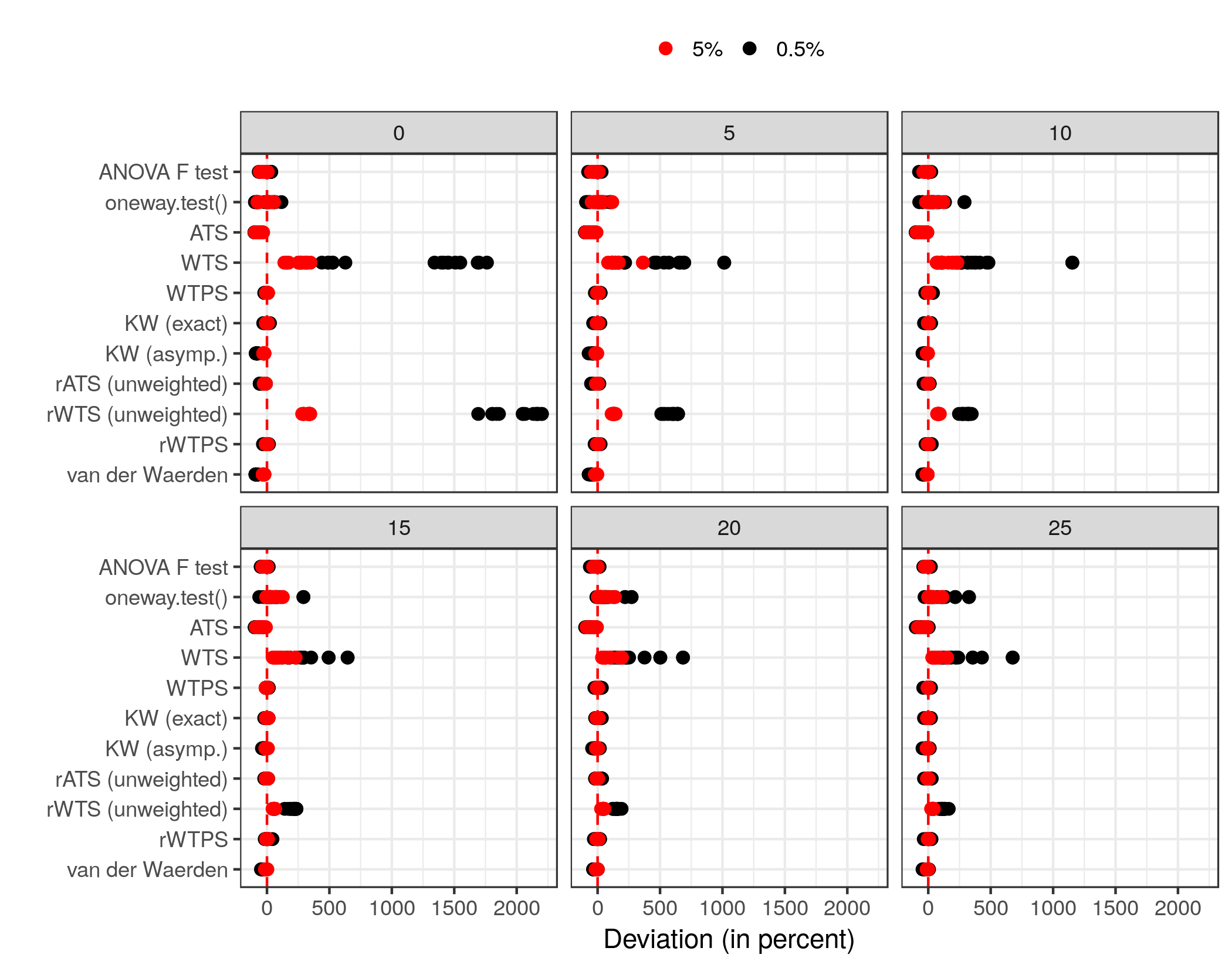}
 \caption{\label{fig:set2_abw} \bf Deviation (in percent) of the prerequired significance levels 5\% (red dots) and 0.5\% (black dots) in the second setting regarding the one-way layout.}
\end{figure}
\newpage
\begin{figure}[!ht]
 \centering
 \includegraphics[width=\textwidth]{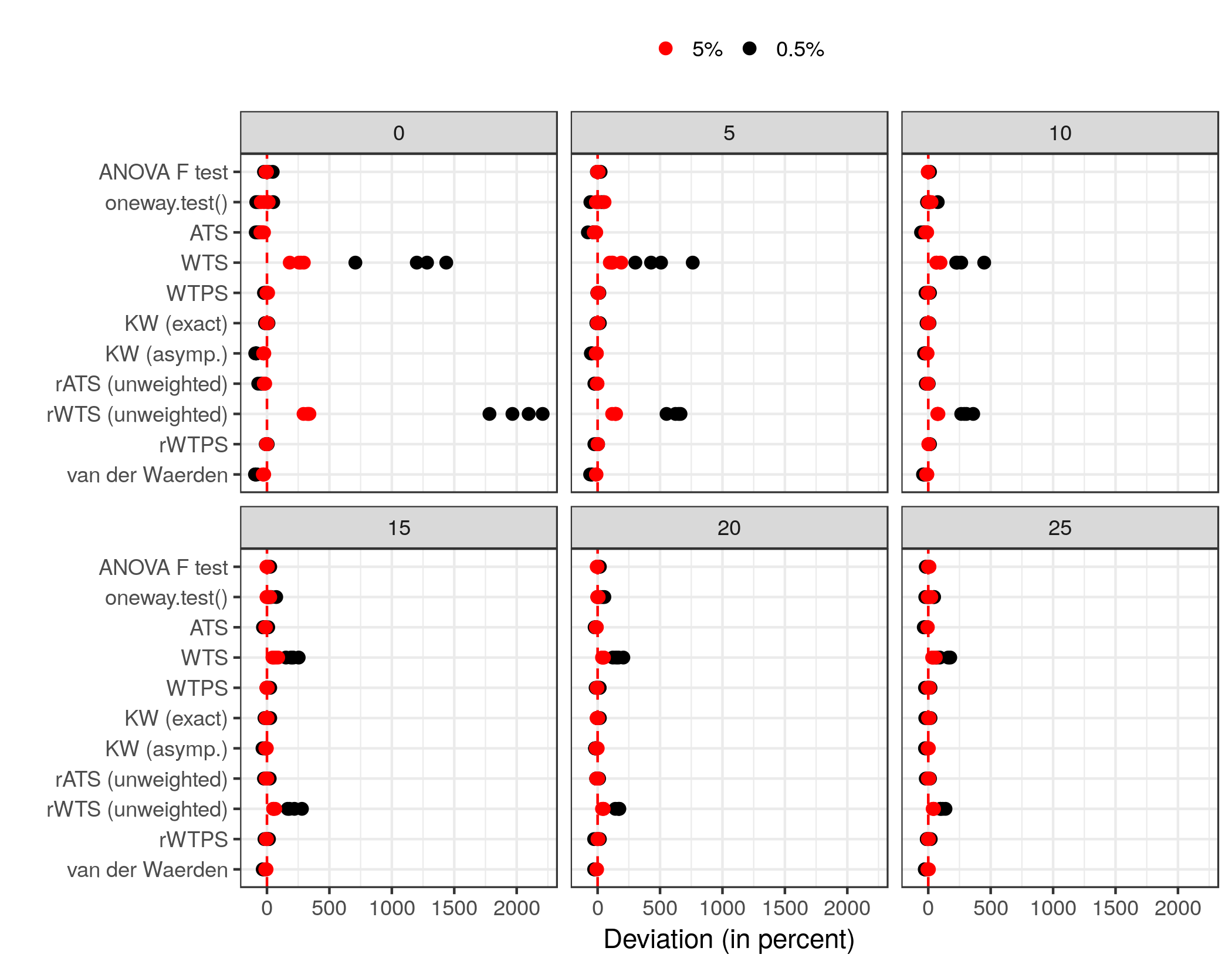}
 \caption{\label{fig:set3_abw} \bf Deviation (in percent) of the prerequired significance levels 5\% (red dots) and 0.5\% (black dots) in the third setting regarding the one-way layout.}
\end{figure}
\newpage
\begin{figure}[!ht]
 \centering
 \includegraphics[width=\textwidth]{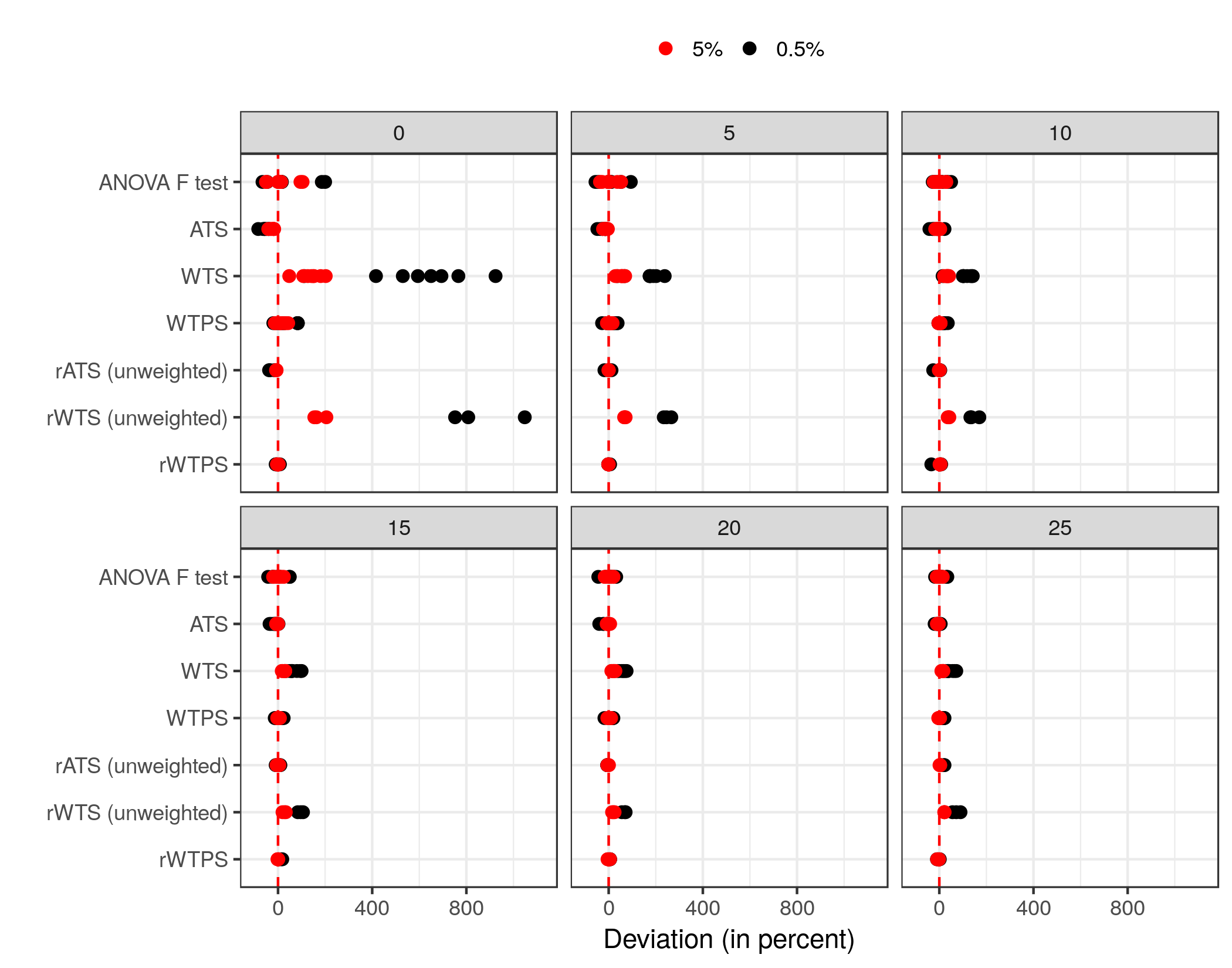}
 \caption{\label{fig:set1_abw_two} \bf Deviation (in percent) of the prerequired significance levels 5\% (red dots) and 0.5\% (black dots) in the first setting regarding the two-way layout.}
\end{figure}
\newpage
\begin{figure}[!ht]
 \centering
 \includegraphics[width=\textwidth]{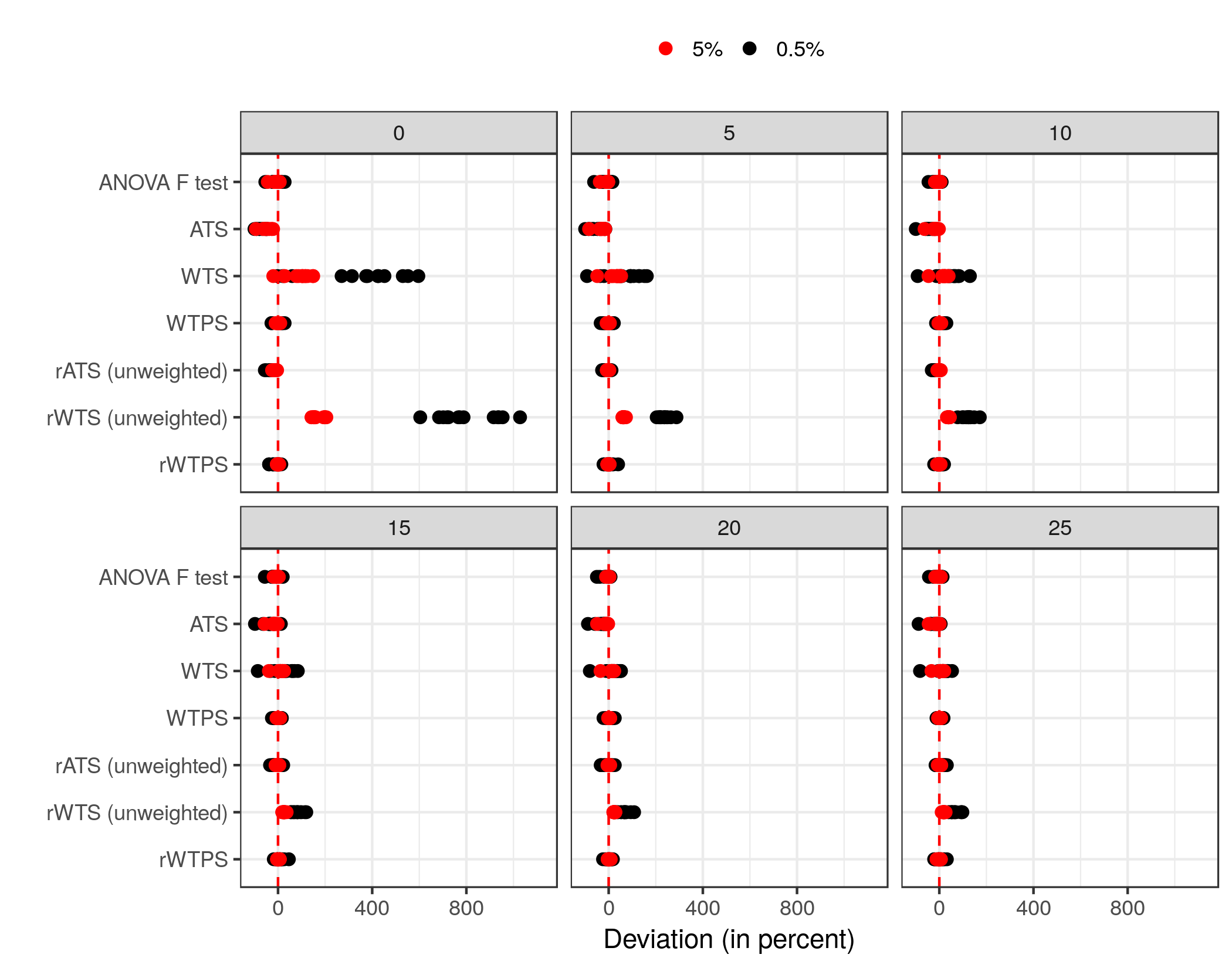}
 \caption{\label{fig:set2_abw_two} \bf Deviation (in percent) of the prerequired significance levels 5\% (red dots) and 0.5\% (black dots) in the second setting regarding the two-way layout.}
\end{figure}
\newpage
\begin{figure}[!ht]
 \centering
 \includegraphics[width=\textwidth]{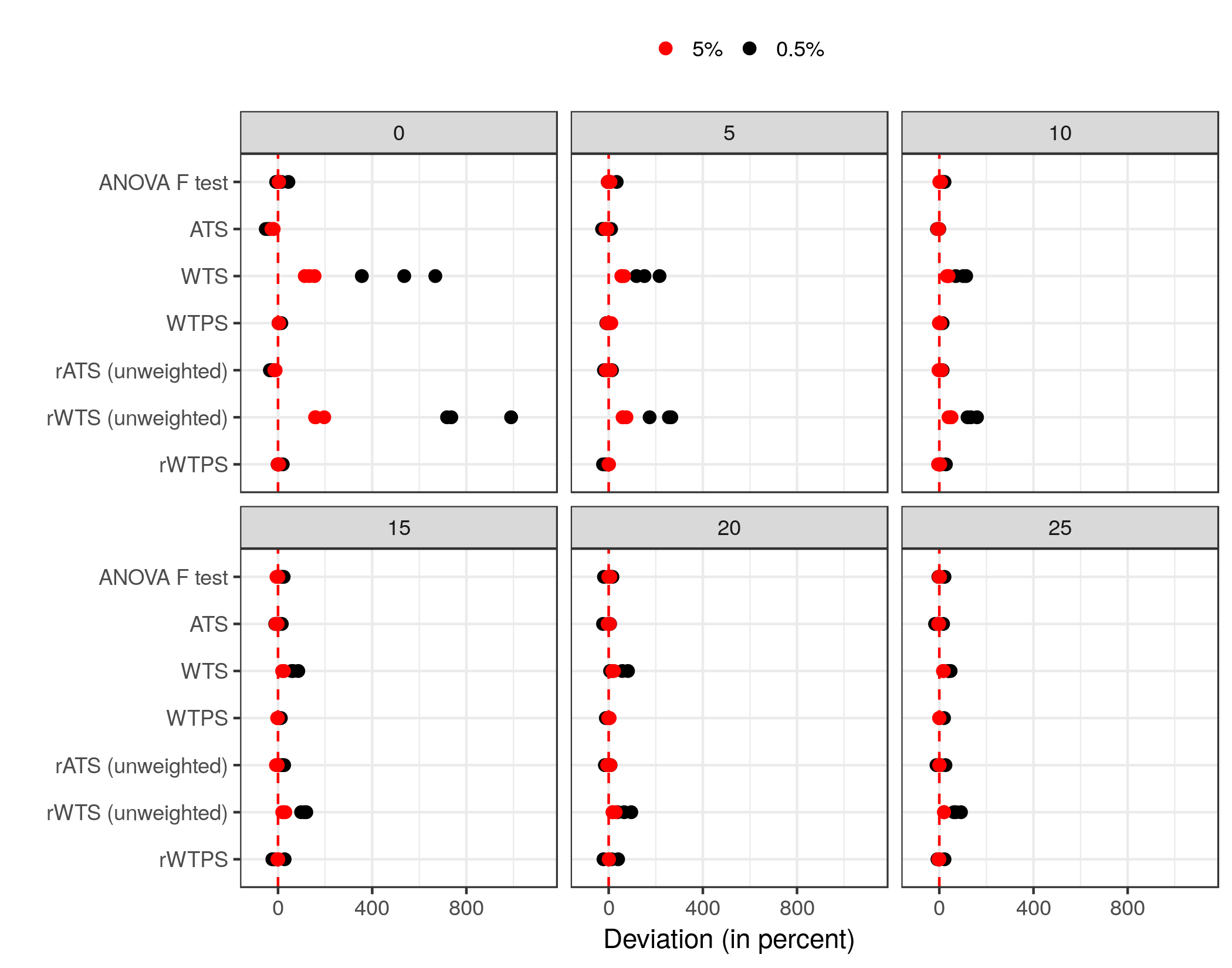}
 \caption{\label{fig:set3_abw_two} \bf Deviation (in percent) of the prerequired significance levels 5\% (red dots) and 0.5\% (black dots) in the third setting regarding the two-way layout.}
\end{figure}

\end{document}